\documentclass[12pt]{article}
\usepackage{epsfig}
\usepackage{wrapfig}
\usepackage{feynmf}
\usepackage{graphicx}
\usepackage{rotating}
\newlength{\dinwidth}
\newlength{\dinmargin}
\begin{document}
\vspace{1 cm}
\newcommand{\Gev}       {\mbox{${\rm GeV}$}}
\newcommand{\Gevsq}     {\mbox{${\rm GeV}^2$} }
\newcommand{\qsd}       {\mbox{${Q^2}$}}
\newcommand{\x}         {\mbox{${\it x}$}}
\newcommand{\smallqsd}  {\mbox{${q^2}$}}
\newcommand{\ra}        {\mbox{$ \rightarrow $}}
\newcommand {\pom}  {I\hspace{-0.2em}P}
\newcommand {\alphapom} {\mbox{$\alpha_{_{\pom}}$}}
\newcommand {\xpom} {\mbox{$x_{_{\pom}}$}}
\newcommand {\xpomp}[1] {\mbox{$x^{#1}_{_{\pom}}\;$}}
\newcommand {\xpoma} {\mbox{$(1/x_{_{\pom}})^a\;$}}
\def\ctr#1{{\it #1}\\\vspace{10pt}}
\def\si{{\rm si}}
\def\Si{{\rm Si}}
\def\Ci{{\rm Ci}}
\def\qsq{Q^{2}}
\def\yjb{y_{_{JB}}}
\def\xjb{x_{_{JB}}}
\def\qjb{\qsq_{_{JB}}}
\def\laproe{\;^{<}_{\sim}\;}
\def\gaproe{\;^{>}_{\sim}\;}
\def\gap{\hspace{0.5cm}}
\renewcommand{\thefootnote}{\arabic{footnote}}
\title {{\normalsize\hfill\texttt{DESY~98-084}\\[3\bigskipamount]}
  \hspace*{-0mm}\large\rm \LARGE Measurement of the Diffractive Cross
  Section in Deep Inelastic Scattering using ZEUS 1994 Data }
\author{ZEUS Collaboration}
\date{ }

\maketitle

\vspace{1 cm}
\begin{abstract}

\noindent The DIS diffractive cross section,
$d\sigma^{di\!f\!f}_{\gamma^* p \to XN}/dM_X$, has been measured in
the mass range $M_X < 15$ GeV for $\gamma^*p$ c.m. energies $60 < W <
200$ GeV and photon virtualities $Q^2 = 7$ to 140 GeV$^2$. For fixed
$Q^2$ and $M_X$, the diffractive cross section rises rapidly with $W$,
$d\sigma^{di\!f\!f}_{\gamma^*p \to XN}(M_X,W,Q^2)/dM_X \propto
W^{a^{diff}}$ with $a^{diff} = 0.507 \pm 0.034
(stat)^{+0.155}_{-0.046}(syst)$ corresponding to a $t$-averaged
pomeron trajectory of $\overline{\alphapom} = 1.127 \pm 0.009
(stat)^{+0.039}_{-0.012} (syst)$ which is larger than
$\overline{\alphapom}$ observed in hadron-hadron scattering. The $W$
dependence of the diffractive cross section is found to be the same as
that of the total cross section for scattering of virtual photons on
protons. The data are consistent with the assumption that the
diffractive structure function $F^{D(3)}_2$ factorizes according to
$\xpom F^{D(3)}_2 (\xpom,\beta,Q^2) = (x_0/ \xpom)^n
F^{D(2)}_2(\beta,Q^2)$. They are also consistent with QCD based models
which incorporate factorization breaking. The rise of $\xpom
F^{D(3)}_2$ with decreasing $\xpom$ and the weak dependence of
$F^{D(2)}_2$ on $Q^2$ suggest a substantial contribution from partonic
interactions.

\end{abstract}

\pagestyle{plain}

\thispagestyle{empty}

\vspace{3 cm}
\clearpage\begingroup
%
%
%
%
\topmargin-1.cm
\evensidemargin-0.3cm
\oddsidemargin-0.3cm
\textheight 680pt
\parindent0.cm
\parskip0.3cm plus0.05cm minus0.05cm
\def\3{\ss}
\pagenumbering{Roman}
                                                   %
\begin{center}
{                      \Large  The ZEUS Collaboration              }
\end{center}
  J.~Breitweg,
  M.~Derrick,
  D.~Krakauer,
  S.~Magill,
  D.~Mikunas,
  B.~Musgrave,
  J.~Repond,
  R.~Stanek,
  R.L.~Talaga,
  R.~Yoshida,
  H.~Zhang  \\
 {\it Argonne National Laboratory, Argonne, IL, USA}~$^{p}$
% ...
\newpage
% ...
$^{\    2}$ now at Univ. of Salerno and INFN Napoli, Italy \\
% ...
$^{  19}$ now at ATLAS Collaboration, Univ. of Munich \\
% ...
$^{  30}$ partially supported by the Foundation for German-Russian Collaboration
DFG-RFBR \\ \hspace*{3.5mm} (grant no. 436 RUS 113/248/3 and no. 436 RUS 113/248/2)\\
% ...
                                                           %
                                                           %
\newpage   
                                                           %
                                                           %
\begin{tabular}[h]{rp{14cm}}
$^{a}$ &  supported by the Natural Sciences and Engineering Research
          Council of Canada (NSERC)  \\
$^{b}$ &  supported by the FCAR of Qu\'ebec, Canada  \\
$^{c}$ &  supported by the German Federal Ministry for Education and
          Science, Research and Technology (BMBF), under contract
          numbers 057BN19P, 057FR19P, 057HH19P, 057HH29P \\
$^{d}$ &  supported by the MINERVA Gesellschaft f\"ur Forschung GmbH,
          the German Israeli Foundation, the U.S.-Israel Binational
          Science Foundation, and by the Israel Ministry of Science \\
$^{e}$ &  supported by the German-Israeli Foundation, the Israel Science
          Foundation, the U.S.-Israel Binational Science Foundation, and by
          the Israel Ministry of Science \\
$^{f}$ &  supported by the Italian National Institute for Nuclear Physics
          (INFN) \\
$^{g}$ &  supported by the Japanese Ministry of Education, Science and
          Culture (the Monbusho) and its grants for Scientific Research \\
$^{h}$ &  supported by the Korean Ministry of Education and Korea Science
          and Engineering Foundation  \\
$^{i}$ &  supported by the Netherlands Foundation for Research on
          Matter (FOM) \\
$^{j}$ &  supported by the Polish State Committee for Scientific
          Research, grant No.~115/E-343/SPUB/P03/002/97, 2P03B10512,
          2P03B10612, 2P03B14212, 2P03B10412 \\
$^{k}$ &  supported by the Polish State Committee for Scientific
          Research (grant No. 2P03B08614) and Foundation for
          Polish-German Collaboration  \\
$^{l}$ &  partially supported by the German Federal Ministry for
          Education and Science, Research and Technology (BMBF)  \\
$^{m}$ &  supported by the Fund for Fundamental Research of Russian Ministry
          for Science and Edu\-cation and by the German Federal Ministry for
          Education and Science, Research and Technology (BMBF) \\
$^{n}$ &  supported by the Spanish Ministry of Education
          and Science through funds provided by CICYT \\
$^{o}$ &  supported by the Particle Physics and
          Astronomy Research Council \\
$^{p}$ &  supported by the US Department of Energy \\
$^{q}$ &  supported by the US National Science Foundation \\
\end{tabular}
                                                           %
\clearpage
\endgroup
\pagenumbering{arabic}

\newpage
\section{Introduction}

Diffraction has been studied extensively in hadron-hadron scattering
at small momentum transfers~\cite{Goulianos}. An elegant
parametrization of the data has been provided by the Regge formalism
through the introduction of a pomeron trajectory~\cite{Gribov} -
~\cite{Donlan1}. The hypothesis that diffraction may have a partonic
component~\cite{Ingelmanschlein} has been substantiated by the
observation of a high transverse energy jet production in diffractive
           $p\overline{p}$ scattering~\cite{UA8}. However, in hadron-hadron
scattering both collision partners are extended objects which makes
the extraction of the underlying partonic process(es) difficult. In
deep-inelastic electron-proton scattering (DIS), on the other hand,
the virtual photon has a pointlike coupling to quarks. The $ep$
collider HERA offers a unique opportunity to study the partonic
structure of diffraction since it gives access to the regime of large
photon virtualities $Q^2$ ($Q^2$ = 10 - 1000 GeV$^2$) and large energy
transfers between the virtual photon and the target proton in its rest
system, $\nu = Q^2/(2m_p x)$ = 2 - 20 TeV, where $x$ is the Bjorken
scaling variable and $m_p$ is the proton mass.

The diffractive dissociation of the virtual photon, first recognized
by the presence of a class of events with a large rapidity
gap~\cite{Zeplrg93,H1eplrg93}, has opened a window for a systematic
study of diffraction in reactions initiated by a hard
probe~\cite{Zeplrgjet93to94} - ~\cite{Zepf2d3lps94}.

In this paper, we present a measurement of the diffractive cross section for
\begin{eqnarray}
 \gamma^*p \to XN
\end{eqnarray}
and of the diffractive structure function
$F^{D(3)}_2$~\cite{Ingelman}. Here, $X$ and $N$ are the particle
systems produced by dissociation of the virtual photon and the
proton. The measurements show that diffraction constitutes a
substantial fraction of the total cross section. The latter is
directly related to the proton structure function $F_2(x,Q^2)$. The
principal signatures for a partonic behaviour in DIS at small $x$ were
found to be a logarithmic dependence of $F_2$ on $Q^2$ associated with
a rapid rise as $x$
decreases~\cite{Zepf293to97,H1epf293to97,Zepf294}. The comparison of
$F^{D(3)}_2$ with $F_2(x,Q^2)$ allows a direct comparison of the QCD
evolution of the two processes with respect to $x$ and $Q^2$.

In QCD, diffraction is characterized by the exchange of a colourless
object, e.g. a colour singlet two-gluon system, between the incoming
virtual photon and proton. The exchange of a colourless system
suppresses QCD radiation, and therefore the production of hadrons, in
comparison with nondiffractive scattering. In the diffractive events
studied in this analysis most of the hadronic energy is carried away
by a low mass nucleonic system $N$ which escapes detection. This
property is used to identify the diffractive contribution. The
diffractive cross section is determined using the $M_X$ method
developed previously to separate the diffractive and nondiffractive
contributions~\cite{Zepsdiff93}.

The present measurement is based on a fivefold larger data sample in a
wider range in $Q^2$ (7 - 140 GeV$^2$) compared to our previous
studies~\cite{Zepf2d393,Zepsdiff93}. The squared momentum transfer $t$
from the virtual photon to the incoming proton is not measured, so the
diffractive contribution was integrated over this variable. The system
$N$ is either a proton or a nucleonic system with mass $M_N < 5.5$
GeV. The new results supersede those presented in~\cite{Zepsdiff93}
which were affected by a technical error simulating the QED radiative
corrections in the Monte Carlo generation used for unfolding. This led
to a steeper energy dependence and a higher intercept of the pomeron
trajectory by about one unit of the quoted error.

\section{Kinematics}

The kinematic quantities used for the description of inclusive DIS,
$e(k) + p(P) \to e(k^{\prime}) + anything$, are $Q^2 = -q^2 =
-(k-k^{\prime})^2$, $x = Q^2/(2P\cdot q)$, $y = (P \cdot q)/(P \cdot
k)$ and $W^2 = Q^2 (1-x)/x + m^2_p \approx Q^2/x$ for $x \ll 1$. Here
$k, k^{\prime}$ are the four-momenta of the initial and final state
positrons; $P$ is the four-momentum of the intial state proton and $y$
is the fractional energy transfer to the proton in its rest frame. For
the range of $Q^2$ and $W$ considered in this paper $W^2 \approx y s$,
where $s = 4E_eE_p$ is the square of the $ep$ c.m. energy, $\sqrt{s} =
300$ GeV. The scaling variables used to describe DIS diffraction are
given by $\xpom =[(P - N)\cdot q]/(P\cdot q) \approx
(M^2_X~+~Q^2)/(W^2~+~Q^2)$ and $\beta = Q^2/[2(P - N)\cdot q] =
x/\xpom \approx Q^2/(M^2_X~+~Q^2)$ where $N$ is the four-momentum of
the outgoing nucleonic system and $M_X$ is the mass of the system into
which the virtual photon dissociated. In models where diffraction is
described by the $t$-channel exchange of a system, for example the
pomeron, $\xpom$ is the momentum fraction of the proton carried by
this system and $\beta$ is the momentum fraction of the struck quark
within this system.

\section{Experimental conditions}

The experimental conditions in 1994 for HERA and the ZEUS detector
were described in our previous paper dealing with the $F_2$
measurement~\cite{Zepf294}. HERA operated with 153 colliding bunches
of 27.5 GeV positrons and 820 GeV protons.  Additional unpaired
positron (15) and proton (17) bunches circulated, which were used to
determine beam related background. The data of this analysis
corresponds to a luminosity of 2.61$\pm$0.04 pb$^{-1}$.

The ZEUS apparatus is described in detail
elsewhere~\cite{Zdetector}. Of particular importance for this analysis
were the uranium-scintillator calorimeter (CAL)~\cite{Zcal}, the
central tracking detector (CTD)~\cite{Zctd}, the small angle rear
tracking detector (SRTD)~\cite{Zsrtd}, the proton remnant tagger
(PRT)~\cite{Zgpdiffmx} and the luminosity monitor (LUMI)~\cite{Zlumi}.

The CAL provides an angular coverage of 99.7\% of $4\pi$ and is
divided into three parts (FCAL, BCAL, RCAL), covering the forward
(proton direction), central and rear regions with pseudorapidity
ranges of $4.3 \le \eta < 1.1, 1.1 \le \eta < -0.75$ and $-0.75 \le
\eta < -3.8$, respectively\footnote{The ZEUS coordinates form a
right-handed system with positive-Z in the proton beam direction and a
horizontal X-axis pointing towards the centre of HERA. The nominal
interaction point is at $X = Y = Z = 0$. The pseudorapidity $\eta$ is
defined as $-\ln (tan \frac{\theta}{2})$, where the polar angle
$\theta$ is taken with respect to the proton beam direction from the
nominal point.}. Each part consists of towers which are longitudinally
subdivided into electromagnetic (EMC) and hadronic (HAC) readout
cells. In test beam measurements, energy resolutions of $\sigma_E/E =
0.18/\sqrt{E}$ for electrons and $\sigma_E = 0.35/\sqrt{E}$ for
hadrons were obtained ($E$ in GeV).

The CTD is a cylindrical drift chamber situated inside a
superconducting solenoid which provides a 1.43 T field. It consists of
72 cylindrical layers covering the polar regions $15^o < \theta <
164^o$ and the radial range 18.2 - 79.4 cm. The transverse momentum
resolution for tracks traversing all CTD layers is $\sigma(p_t)/p_t
\approx \sqrt{(0.005 p_t)^2 + (0.016)^2}$, with $p_t$ in GeV. The
vertex position of a typical multiparticle event is determined from
the tracks to an accuracy of typically $\pm 1$ mm in the $X,Y$ plane
and $\pm 4$ mm in $Z$.

The PRT is used to tag diffractive events where the proton
dissociated.  The PRT consists of two layers of scintillation counters
installed perpendicular to the beam at $Z = 5.15$ m, i.e. downstream
of FCAL and beam collimator C4. The two layers are separated by a 2 mm
thick lead absorber. Each layer is split into two halves with two
counters each which are independently read out by
photomultipliers. The counters have an active area of dimensions $30
\times 26$ cm$^2$ with a hole of $6.0 \times 4.5$ cm$^2$ at the centre
to accomodate the HERA beam pipe. The pseudorapidity covered by the
PRT is $4 \laproe \eta \laproe 6$. The data with useful PRT
information correspond to an integrated luminosity of $0.7$ pb$^{-1}$.

\section{Reconstruction of kinematic variables}

The kinematic variables $x,Q^2,W$ and $M_X$ were determined from
calorimeter and tracking information. The calorimeter cells were
required to have energy deposits above 60 MeV in the EMC section and
110 MeV in the HAC section unless these energy deposits were isolated
in which case thresholds of 120 MeV (160 MeV) in the EMC (HAC)
sections were used. An energy-momentum vector
$(E_j,p_{Xj},p_{Yj},p_{Zj})$ with $E^2_j = p^2_{Xj}+p^2_{Yj}+p^2_{Zj}$
was assigned to every calorimeter cell $j$ with energy deposition
$E_j$. The cell angles were calculated from the geometric centre of
the cell and the vertex position of the event. The algorithm used to
identify the scattered positron was based on a neural
network~\cite{Zsira} which included information from the CAL. The
systematic uncertainty in the energy determination of the scattered
positron is 2\% at 10 GeV decreasing linearly to 1\% at 27.5 GeV for
the $Q^2$ region considered in this analysis. The momenta of the
particles of the system $X$ were reconstructed from clusters found in
the calorimeter and from tracks in the
CTD~\cite{Zepf2d3lps94,Briskinthesis}. The inclusion of tracking
information improves the $M_X$ resolution and reduces the sensitivity
to the losses due to inactive material in front of the calorimeter. A
systematic uncertainty of 3\% is assigned to the hadronic energy
measurement. The selected clusters and tracks are called energy flow
objects (EFO's).

The kinematic variables $x$, $Q^2$ and $W$ were determined with the
double angle (DA) method~\cite{Bentvelsen} in which only the angles of
the scattered positron ($\theta^{\prime}_e$) and the produced hadronic
system ($\gamma_H$) are used. The angles were determined from the
EFO's.  In the DA method, in order that $\gamma_H$ be well measured,
it is necessary to require a minimum of hadronic activity in the
calorimeter away from the forward direction. A suitable quantity for
this purpose is the hadronic estimator of the variable
$y$~\cite{Jacquet}, defined by $y_{JB} = \sum_j \left(
E_j-p_{Zj}\right)/2E_{e}$, where the sum runs over all EFO's not
assigned to the scattered positron.

We study events of the type $ep \to e + X + rest$, where $X$ denotes
the hadronic system observed in the central detector (CAL and CTD) and
$rest$ the particle system escaping detection through the beam holes.
The mass $M_X$ of the system $X$ was determined by summing over all
EFO's not assigned to the scattered positron:
\begin{eqnarray}
(M_X^{meas})^2 = (\sum_j E_j)^{2}-(\sum_j p_{Xj})^{2}-(\sum_j
p_{Yj})^{2}-(\sum_j p_{Zj})^{2} \; \; \; .
\label{mass}
\end{eqnarray}

\section{Trigger and event selection}

The event selection at the trigger level was identical to that used
for the $F_2$ analysis of the same data~\cite{Zepf294}. The off-line
cuts were also similar to those applied previously. The energy of the
scattered positron had to satisfy $E^{\prime}_e >$ 10 GeV to ensure
reliable positron finding and to suppress the photoproduction
background. The variable $y$, calculated from the scattered positron,
was required to satisfy $y_e < 0.95$ to suppress events with spurious
low energy positrons. The impact point of the positron on the face of
the RCAL had to lie outside a square of side 26 cm centered on the
beam axis (boxcut) to ensure full containment of the positron
shower. The requirement $y_{JB} >$ 0.02 ensured a good measurement of
the angle $\gamma_H$ and hence of $x$. By requiring 40 $< \delta <$ 70
GeV, where $\delta = \sum_j (E_j-p_{Zj})$ summed over all EFO's,
including those belonging to the scattered positron, both the
background from photoproduction and the radiative corrections were
reduced. The primary event vertex was determined from the tracks. If
no tracking information was present (9.2\% of the events) the vertex
position was set to the nominal interaction point.

After the selection cuts and removal of events from QED Compton
scattering and cosmic rays, the DIS sample contained 304k events. For
the present analysis, 157k events with $60 < W < 200$ GeV, $7 < Q^2 <
140$ GeV$^2$ were used.

\section{Monte Carlo simulation}

Monte Carlo simulations were used for testing the validity of the
subtraction of the nondiffractive contribution, for understanding the
contribution from double dissociation ($\gamma^*p \to XN$), for
unfolding the produced event distributions from the measured ones, for
determining the acceptance and for estimating the systematic
uncertainties. The detector simulation is based on the GEANT
program~\cite{geant} and incorporates our present understanding of the
detector and the trigger and test beam results.

Hadronic final states from diffractive DIS interactions where the
proton does not dissociate, $ep \to e X p$, were modelled with
RAPGAP~\cite{rapgap,Briskinthesis} modified to include low-mass vector
meson production.  RAPGAP is based on a factorizable
model~\cite{Ingelmanschlein} in which the incoming proton emits a
pomeron. The interaction of the virtual photon with this pomeron is
described by an effective structure function $F_2^{\pom}(\beta,Q^2)$
that is independent of the process of emission of the pomeron and
where the partons of the pomeron take part in the hard scattering. The
parton densities of the pomeron are evolved from a starting scale
$Q^2_0 = 4$ GeV$^2$ using the next-to-leading order DGLAP
equations~\cite{dglap}.  The free parameters are adjusted to reproduce
the results on the diffractive structure function $F_2^{D(3)}$
measured by H1~\cite{H1epf2d394}. The momentum sum rule was not
imposed. RAPGAP was used with the parton showering scheme of ARIADNE
4.08~\cite{ariadne}, which is based on the color dipole model and
includes the first order QCD matrix elements, and the Lund
fragmentation scheme~\cite{lund} as implemented in JETSET
7.4~\cite{jetset}. The region of low masses (below 1.1 GeV) is tuned
to the measured ratio of $\rho:\phi$ resonance
production~\cite{Zeprho,Zepphi,H1epv} and allowing for a contribution
from $\omega$ production.

The contribution from the diffractive process where the proton
dissociates, $\gamma^* p \to X N^{dissoc}$, was simulated using
EPSOFT~\cite{Kasprzak}.  Assuming factorisation and a Triple Regge
formalism ~\cite{Mueller,Fiefox} for modelling $\gamma^* p \to X p$
and $\gamma^* p \to X N^{dissoc}$, the measured cross sections for
elastic and single diffractive dissociation in $pp$ (and
$\overline{p}p$) scattering, $pp \to p p$ and $pp \to p N^{dissoc}$,
were used to relate $\gamma^* p \to X N^{dissoc}$ to $\gamma^* p \to X
p$.

For testing the procedure used to separate the diffractive and
nondiffractive contributions, events from standard nondiffractive DIS
processes were produced within the framework provided by DJANGO
6.0~\cite{django}. First order electroweak corrections were generated
with HERACLES 4.5~\cite{herac}. The positron-proton kinematics was
produced with LEPTO 6.5~\cite{lepto} which was interfaced to ARIADNE
4.08 for the simulation of the parton shower process. For
fragmentation JETSET 7.4 was used. The parton densities of the proton
were chosen to be the MRSA set~\cite{Martin}.

All Monte Carlo events were passed through the standard ZEUS detector and
trigger simulations as well as the event reconstruction package.

\section{Precision of kinematic variables and binning}
The resolutions expected for the kinematical variables were estimated
from Monte Carlo (MC) simulation. The intervals in $W$ were chosen
with equidistant bins in $\ln W^2$, thereby providing approximately
equal numbers of events in each $W$ bin. Here, and in the following,
masses and energies are given in units of GeV. The average
r.m.s. resolution $\sigma (\ln W^2)$ is 0.23 in the lowest ($W,Q^2$)
bin decreasing to 0.09 in the highest ($W,Q^2$) bin. A bin width of
$\Delta \ln W^2 = 0.4$ was chosen. For $Q^2$ the r.m.s. resolution is
less than 1 GeV$^2$ in the lowest $Q^2$ bin increasing to 3~GeV$^2$ in
the highest $Q^2$ interval.

The mass $M_X^{meas}$, determined from the EFO's, has to be corrected
for energy losses in the material in front of the calorimeter and for
acceptance. The correction was determined by comparing, for Monte
Carlo (MC) generated events, the MC measured mass, $M_X^{MCmeas}$, to
the generated mass, $M_X^{MCgen}$, of the system $X$.  The resulting
corrections to determine the diffractive cross section were performed
in three steps.

In the first step an overall mass correction factor was calculated
from the ratios of measured to generated masses, $ f(M_X^{MCmeas}) =
M_X^{MCmeas}/M_X^{MCgen}$, as a function of $M_X$, $W$ and $Q^2$. The
variation of $f(M_X^{MCmeas})$ with $M_X, W, Q^2$ was found to be
sufficiently small ($\le 6\%$) for the $M_X$ range used in this
analysis, $1.5 < M_X < 15$~GeV, so that it could be neglected in the
first step of the mass correction.  The average correction factor was
$\overline{f} = 0.80$. The same correction factor was used for masses
below 1.5~GeV.  The correction factor $\overline{f}$ was applied to
obtain the corrected mass value,
$M_X^{cor}=M_X^{meas}/\overline{f}$. The r.m.s resolution for $M_X$
was $\sigma(M_X)/\sqrt{M_X} \approx 60\%$GeV$^{\frac{1}{2}}$ on
average. All $M_X$ distributions presented below refer to $M^{cor}_X$.

For diffractive production, a comparison of the MC-generated
distributions with the MC-measured distributions show a depletion of
events at the high mass end and an excess at somewhat lower mass
values. This (small) distortion $r \equiv N^{MCmeas}/N^{MCgen}$ is
caused mainly by particles lost through the forward beam hole. Since
$r$ was found to be independent of $Q^2$, within errors, it was
determined in bins of $\ln M^2_X$ for the different $W$ intervals
averaged over $Q^2$ yielding $\overline{r}(\ln M^2_X,W)$; no smoothing
was applied to $\overline{r}$. The distortion was taken into account
in extracting the diffractive contribution.

In the final step, the diffractive cross section was determined by an
unfolding procedure discussed below taking into account, for each
($M_X, W, Q^2$) interval, the proper mass correction as determined
from the MC simulation.

Results are presented for the intervals in $W$: 60 - 74, 74 - 90, 90 -
110, 110 - 134, 134 - 164, 164 - 200 GeV; in $Q^2$: 7 - 10, 10 - 20,
20 - 40, 40 - 140 GeV$^2$, with average $Q^2$ values of 8, 14, 27, 60
GeV$^2$; in $M_X$: $<$ 3, 3 - 7.5, 7.5 - 15 GeV, with average $M_X$
values of 2, 5 and 11 GeV.

\section{Characteristics of the $M_X$ distributions}

\begin{figure}[ht]
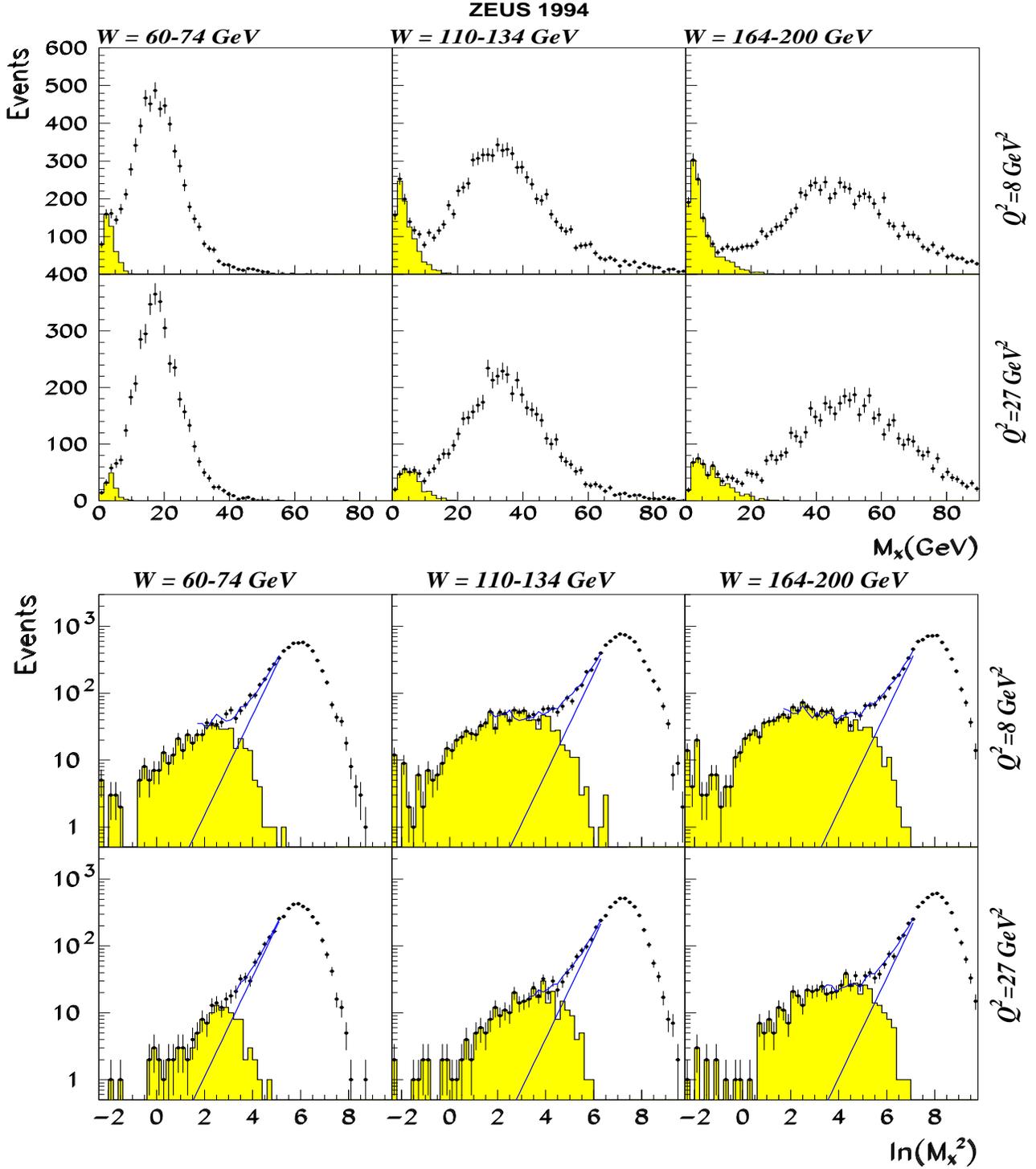

\begin{center}
\epsfig{file=Fig.1a,width=17.5cm,height=10cm}
\epsfig{file=Fig.1b,bburx=798pt,bbury=510pt,width=17.5cm,height=10cm}
\end{center}
\caption[]{Reaction $\gamma^* p \to X + anything$, where $X$ is the system
         observed in the detector.  Top: Distributions of $M_X$, the
         corrected mass of the system $X$. The distributions are not
         corrected for acceptance effects.  The shaded histograms show
         the distributions of events with $\eta_{max}<1.5$.\newline
         Bottom: Same distributions as above presented in terms of
         $\ln M^2_X$.  The straight lines give the nondiffractive
         contributions as obtained from the fits.  The upper curves
         show the fit results for the sum of the diffractive and
         nondiffractive contributions.}
\label{f:mass}
\end{figure}

The method used to separate the diffractive and nondiffractive
contributions is based on their very different $M_X$ distributions.
The mass distributions for typical ($W,Q^2$) intervals are presented
in Fig.~\ref{f:mass}(top) in terms of $M_X$.  The distributions shown
were not corrected for acceptance. Two distinct groups of events are
observed, one peaking at low $M_X$ values, the other at high $M_X$
values.  While the position of the low mass peak is independent of
$W$, the high mass peak moves to higher values as $W$ increases.

Most of the events in the low mass peak exhibit the large rapidity gap
that is characteristic of diffractive production. This may be seen
from the shaded areas in Fig.~\ref{f:mass} which show the
distributions of events with $\eta_{max}<1.5$ where $\eta_{max}$ is
the largest pseudorapidity at which energy deposition in the
calorimeter or a track was observed. This value corresponds to a
pseudorapidity gap in the detector larger than $\simeq 2.5$ units
since the beam hole of the FCAL is at $\eta_{edge} \simeq 4$.

In Fig.~\ref{f:mass}(bottom) the mass distributions are presented in
terms of $\ln M^2_X$. In this representation, the $M_X$ distributions
exhibit a simple scaling behaviour. The low mass peak shows up as a
plateau-like structure at low $\ln M_X^2$, most notably at high $W$
values.  The shaded histograms show again the distributions of the
events with a large rapidity gap, $\eta_{max} < 1.5$, which account
for most of the observed plateau~\footnote{Note that a cut on
$\eta_{max}$ will select events produced by colourless exchange, from
both diffraction and from reggeon exchange but does not, in general,
allow the extraction of the full diffractive contribution (see
also~\cite{Ellisros}).}. The high mass peak has a steep exponential
fall-off, $d{\cal N}/d\ln M^2_X \propto \exp (b\ln M^2_X)$, towards
smaller $\ln M_X^2$ values. The peak position of the high mass events
changes proportionally to $\ln W^2$, i.e. shows scaling in $\ln
(M^2_X/W^2$) and the slope, $b$, of the exponential in $\ln M_X^2$ is
approximately independent of $W$ and $Q^2$. These characteristics are
properties of events with uniform, random and uncorrelated particle
production along the rapidity axis where particles are accepted in a
limited range of rapidity. In models such as the Feynman gas model or
one dominated by longitudinal phase space~\cite{Feynman}, the slope
$b$ represents the particle multiplicity per unit of
rapidity.~\footnote{The pseudorapidity gap $\Delta \eta$ is related to
$\ln M^2_X$ via $\Delta \eta \approx \ln (W^2/M^2_X)$ as discussed in
detail in~\cite{Zepsdiff93}.} The exponential in $\ln M^2_X$ and the
scaling in $\ln (M^2_X/W^2)$ are directly connected to the exponential
suppression of large rapidity gaps by QCD radiation. The latter
populates the region between the struck quark and the coloured proton
remnant.

\begin{figure}[ht]
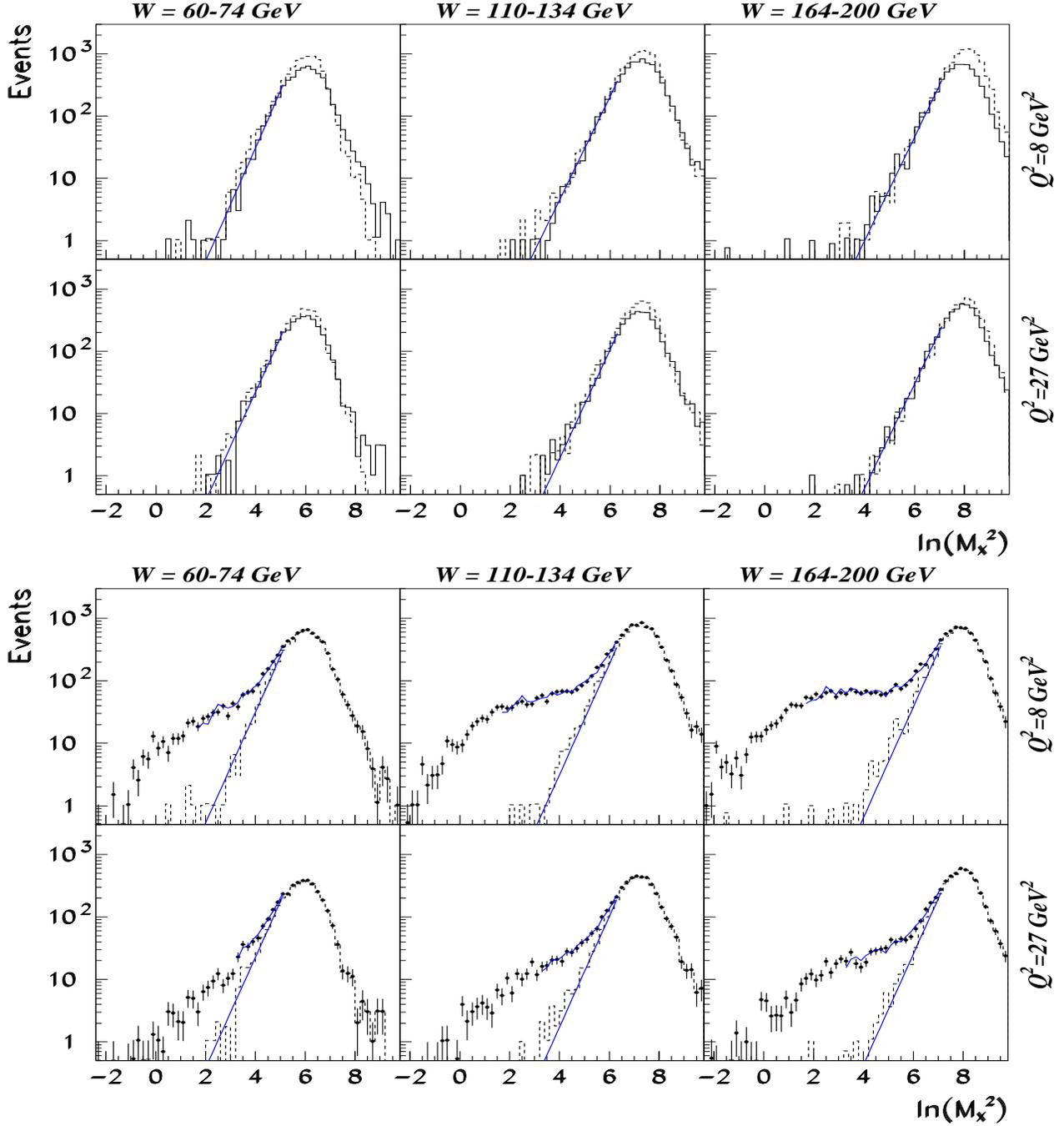

\begin{center}
\epsfig{file=Fig.2a,width=17.5cm,height=10cm}
\epsfig{file=Fig.2b,bburx=798pt,bbury=510pt,width=17.5cm,height=9cm}
\end{center}
\caption[]{Reaction $\gamma^* p \to X + anything$, where $X$ is the system
         observed in the detector.  Top: Distributions in $\ln M^2_X$
         as predicted by ARIADNE for the nondiffractive contribution
         at the generator level (solid histograms) and detector level
         (dashed histograms) for the $W$ intervals and $Q^2$ values
         indicated. The straight lines show the results of the fits to
         the distributions at the detector level.  Bottom:
         Distributions in $\ln M^2_X$ for the sum of the diffractive
         and nondiffractive contributions as predicted at the detector
         level by RAPGAP plus ARIADNE (points with error bars) and for
         the nondiffractive contribution alone (dashed
         histograms). The straight lines show the results for the
         nondiffractive contribution obtained from fitting the sum of
         the diffractive and nondiffractive contributions with
         $b^{MC}_{nom}$. The upper curves show the fit results for the
         sum of the diffractive and nondiffractive contributions.}
\label{f:lnmcdmfit6}
\end{figure}

These characteristics are also properties of realistic models for
particle production in deep inelastic scattering. ARIADNE, which gives
a good description of particle production by DIS at HERA, also
exhibits a pure exponential fall-off with $\ln M^2_X$ and scaling in
$\ln M^2_X/W^2$.This is shown in Fig.~\ref{f:lnmcdmfit6}(top) which
presents the MC simulation of the nondiffractive contribution at the
generator level as a function of $\ln (M^2_X)$ for different ($W,Q^2$)
intervals (solid histograms) where only particles generated with $\eta
< 4.3$ were included. The slope $b$ is independent of $W$ and $Q^2$,
which is in agreement with the experimental observation that, for DIS
events, the average charged particle multiplicity per unit of
pseudorapidity $\eta$ is independent of $W$ and
$Q^2$~\cite{Pavel,Kuhlen}.

The comparison of the distributions at the generator level (dotted
histograms) and the detector level (dashed histograms) in
Fig.~\ref{f:lnmcdmfit6}(top) shows that the exponential fall-off with
$\ln M^2_X$ is not affected by detector effects.

\section{Extraction of the diffractive contribution}
\label{diffextraction}

In diffractive events, the system $X$ resulting from the dissociation
of the virtual photon is almost fully contained in the detector while
the outgoing proton or low mass nucleonic system escapes through the
forward beam hole. Furthermore, diffractive dissociation prefers small
$M_X$ values and leads to an event distribution of the form $d{\cal
N}/dM^2_X \propto 1/(M^2_X)^{(1+n)}$ corresponding to $d{\cal N}/d \ln
M^2_X \propto 1/(M^2_X)^n$, approximately independent of $W$. At high
energies and for large $M_X$, one expects $n \approx 0$, leading to a
constant distribution in $\ln M^2_X$. Such a mass dependence is seen
in diffractive dissociation of $pp$ and $p\overline{p}$
scattering~\cite{Goulianos,CDF}. A value of $n \approx 0$ is also
expected in some models~\cite{Nikzak} - ~\cite{Donlan2} for
diffractive DIS as a limiting value for the fall-off of the mass
distribution.

We identify the diffractive contribution as the excess of events at
small $M_X$ above the exponential fall of the nondiffractive
contribution in $\ln M^2_X$. This will be referred to as the $M_X$
method for the determination of the diffractive component.  The
exponential fall permits the subtraction of the nondiffractive
contribution and, therefore, the extraction of the diffractive
contribution without assuming the precise $M_X$ dependence of the
latter. The distribution is expected to be of the form:
\begin{eqnarray}
\frac{d{\cal N}}{d\,\ln M_X^2} = D +  c\, \exp (b\,\ln {M_X^2}), ~~~~
{\rm for} ~~~ \ln M_X^2 \leq \ln W^2 -\eta_0.
\label{eq:diffnondiff}
\end{eqnarray}
Here, $D$ denotes the diffractive contribution and the second term the
nondiffractive contribution. The diffractive term $D$ is multiplied by
the distortion function $\overline{r}(\ln M^2_X,W)$ discussed
above. The quantity ($\ln W^2 -\eta_0$) specifies the maximum value of
$\ln M^2_X$ up to which the exponential behaviour of the
nondiffractive part holds. A value of $\eta_0 = 3.0$ was found from
the data~\cite{Zeplrg93}. We apply Eq.~\ref{eq:diffnondiff} in a
limited range of $\ln M^2_X$ to determine the parameters $b$ and $c$
of the nondiffractive contribution. The diffractive contribution is
not taken from the fit result for $D$ but is determined by subtracting
from the observed number of events the nondiffractive contribution
found from the fit values of $b$ and $c$.

The diffractive contribution in various models of
diffraction~\cite{Nikzak,pompyt,Donlan2} is expected to be a slowly
varying function of $\ln M^2_X$ when $M^2_X > Q^2$ and to approach, in
the asymptotic limit, a constant $\ln M^2_X$ distribution at large
$M_X$, $D = constant$. The nondiffractive contribution in the
($M_X,W,Q^2$) bins, was determined in two steps. This procedure will
be referred to as the nominal analysis. In the first step, the slope
$b$ was determined as an average over the fits to the data of the high
$W$ intervals $134 - 164, 164 - 200$ GeV for $Q^2 = 7 - 10$ and $10 -
20$ GeV$^2$ in the restricted $M_X$ range, $\ln 10 Q^2 < \ln M^2_X <
\ln W^2 - \eta_0$. The fits yielded $b = b_{nom} = 1.72\pm 0.07$. This
value is smaller than that predicted with ARIADNE by about $10\%$ (see
below).

In the second step, the fits were repeated for all ($W,Q^2$) intervals
in the wide $M_X$ range, $\ln Q^2 < \ln M^2_X < \ln W^2 - \eta_0$,
using $b = b_{nom}$ as fixed parameter and assuming $D =
constant$. The fit results for the nondiffractive contribution are
shown in Fig.~\ref{f:mass} by the straight lines, those for the sum of
the diffractive and nondiffractive distributions by the upper curves;
the latter fluctuate since no smoothing was applied for the distortion
correction $\overline{r}$. The fit results for the sum of the
diffractive and nondiffractive contributions give a good description
of the measured event distributions. Subtraction of the nondiffractive
contribution yielded the number of diffractive events. From these, the
diffractive cross section was obtained by unfolding. Fits were also
performed with different forms for $D$. These were used for estimating
the corresponding systematic uncertainties (see below).

The fit procedure for separating the diffractive from the
nondiffractive contribution was tested with MC simulated event samples
from diffractive (using RAPGAP) and nondiffractive scattering (using
ARIADNE).  It was first checked that the slope of the exponential
fall-off for the nondiffractive contribution is not affected by
detector effects. The $\ln M^2_X$ distributions predicted for the
nondiffractive contribution are shown in Fig.~\ref{f:lnmcdmfit6}(top)
at the generator (dashed histograms) and detector (solid histograms)
levels for the $W$ intervals and $Q^2$ values. The straight lines show
the fits to the MC measured distributions. The value of the slope
$b^{MC gen}$ at the generator level (average value $b^{MCgen} =
1.97\pm0.03$) agrees well with the value $b^{MC meas}$ at the detector
level (average value $b^{MC meas} = 1.96\pm0.03$) for each ($W,Q^2$)
interval. Fits performed to the sum of the diffractive and
nondiffractive contributions in the restricted $M_X$ range are shown
in Fig.~\ref{f:lnmcdmfit6}(bottom). They resulted in an average slope,
$b^{MC}_{nom} = 2.04 \pm 0.09$, which agrees well with the $b$-values
found from the fits to the nondiffractive sample alone. In this figure
the distributions of $\ln M^2_X$ are displayed for the sum of the
diffractive and nondiffractive contributions as predicted at the
detector level (points with error bars) and for the nondiffractive
contribution alone (dashed histograms). In the next step, the sum of
the diffractive and nondiffractive contributions were fitted in the
wide $M_X$ range using $b = b^{MC}_{nom}$. The straight lines show the
nondiffractive contribution as obtained from the fits; they give a
good description of the input distributions for the nondiffractive
contribution shown as the dashed histograms. The upper curves show the
fit results for the sum of the diffractive and nondiffractive
contributions. A good description of the input distributions is
observed. Using the fits with $b = b^{MC}_{nom}$, the nondiffractive
contribution was subtracted which yielded, for every ($W,Q^2$)
interval, the number of MC measured diffractive events. From this
result, the number of MC produced events was determined by
unfolding. Comparison with the generated (i.e. input) numbers of
diffractive events showed very good agreement for all ($M_X,W,Q^2$)
intervals.

Before applying the fitting and unfolding procedure, the number of
events observed in the ($M_X,W,Q^2$) intervals were corrected for the
contribution from positron (proton) beam gas scattering. Averaged over
all events, the contamination from positron (proton)-gas scattering
amounted to 2.1$\%$ (1.0$\%$). For the nominal analysis, which did not
require an event vertex, beam-gas contributions were small for $M_X >
3$ GeV and negligible for $Q^2 $ above 20 GeV$^2$. Sizeable
contributions were observed for $M_X < 3$ GeV when $W > 90$ GeV and
$Q^2 < 20$ GeV$^2$ where they amounted to 14\% on average for
positron-gas scattering and $\le 10\%$ for proton-gas scattering. To
estimate the systematic uncertainties the analysis was also performed
requiring an event vertex (see below). In this case the background
from beam-gas scattering was negligible everywhere.

The number of diffractive events produced, ${\cal
N}^{di\!f\!f}_{prod}$, was obtained from the number of events
determined from the fits, ${\cal N}^{di\!f\!f}_{meas}$, by an inverse
matrix unfolding procedure which took into account detector effects
such as bin-to-bin migration, trigger biases and event selection
cuts. It was checked that the diffractive model (RAPGAP) describes the
energy flow as a function of $\eta$ for $\eta_{max} < 1.5$ for all
($M_X,W,Q^2$) bins~\cite{Briskinthesis} as well as the observed $M_X$
distributions in the region $M_X < 3$ GeV.

\subsection{Contribution from nucleon dissociation}

An estimate of the contribution from diffractive dissociation of the
proton, $\gamma^* p \to XN^{dissoc}$, (double dissociation) to the
diffractive cross section presented below was obtained by comparing
the contributions for $\gamma^* p \to Xp$ (single dissociation) with
an identified proton in the leading proton spectrometer
LPS~\cite{Zepf2d3lps94} and for $\gamma^* p \to XN$ determined in this
analysis. This led to the fractional contribution from double
dissociation $(XN - Xp)/XN = XN^{dissoc}/XN = (31 \pm 15) \%$.

\begin{figure}[ht]
\begin{center}
\epsfig{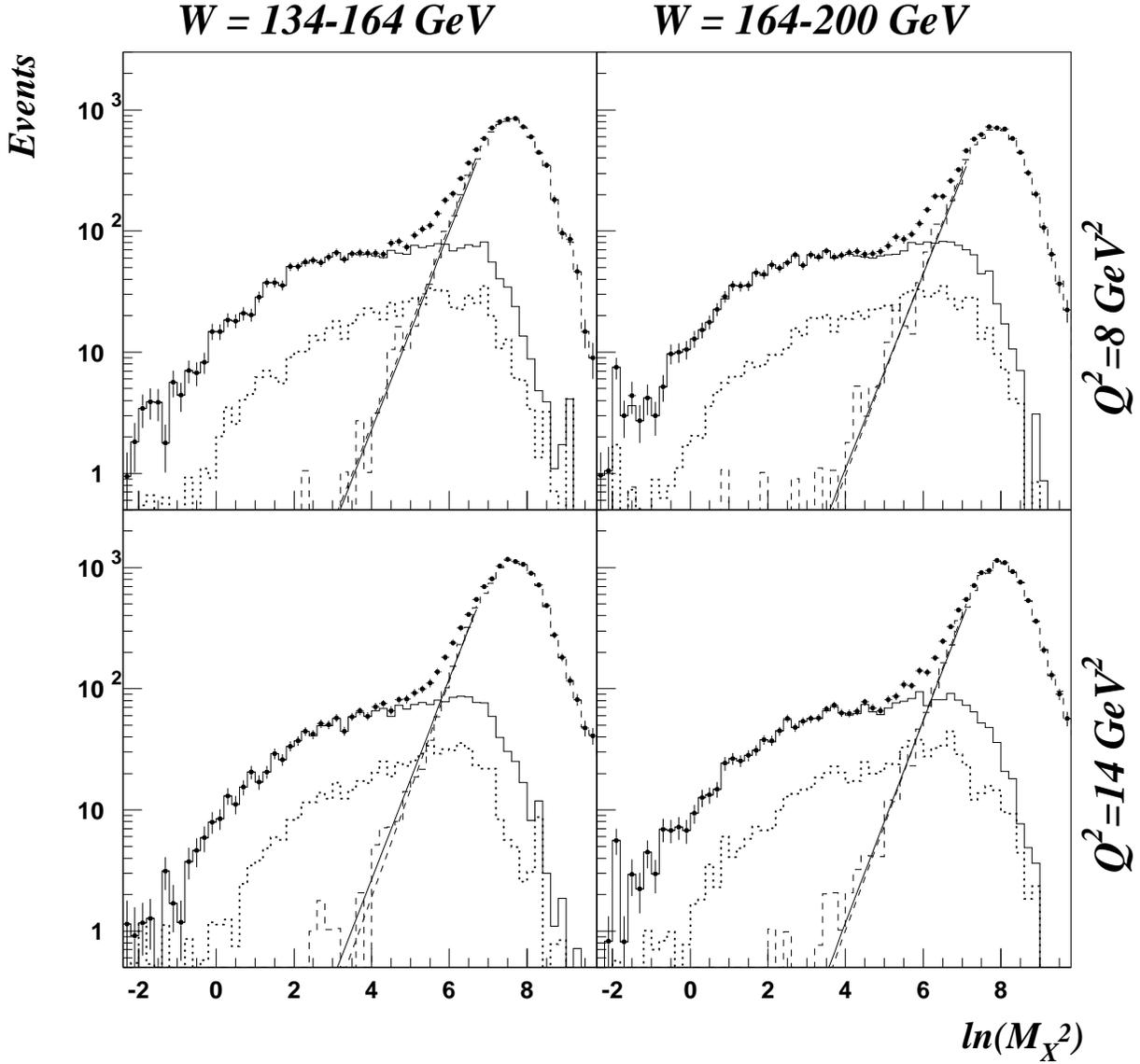}
\end{center}
\caption[]{Reaction $\gamma^* p \to X + anything$, where $X$ is the system
         observed in the detector.  Shown are distributions of $\ln
         M^2_X$ at the detector level.  The dotted histograms show the
         diffractive contributions from $\gamma^* p \to XN^{dissoc}$
         as predicted by EPSOFT.  The solid histograms show the sum of
         the diffractive contributions from $\gamma^* p \to
         XN^{dissoc}$ and $\gamma^* p \to Xp$ (the latter as predicted
         by RAPGAP).  The dashed histograms show the nondiffractive
         contributions as predicted by ARIADNE.  The points with error
         bars show the sum of the diffractive and the nondiffractive
         contributions. The dashed straight lines show the fits
         performed to the nondiffractive contributions alone.  The
         straight lines show the results for the nondiffractive
         contribution from fitting the sum of the diffractive and
         nondiffractive contributions with $b^{MC(pN)}$.}
\label{f:lncdmrapepsoft}
\end{figure}

The procedure for extraction of the diffractive contribution was
tested at the MC level also for the case where an additional
contribution from diffractive dissociation of the proton was
present. The events for $\gamma^* p \to XN^{dissoc}$ were simulated
using EPSOFT. In Fig.~\ref{f:lncdmrapepsoft} the $\ln M^2_X$
distributions are presented at the detector level for the diffractive
contributions from $\gamma^* p \to XN^{dissoc}$ (dotted histograms)
and for the sum of $\gamma^* p \to Xp\; + \; XN^{dissoc}$ (solid
histograms). Also given are the nondiffractive contributions (dashed
histograms) and the sum of the diffractive and nondiffractive
contributions (points with error bars). Fits to the sum of the
diffractive and nondiffractive contributions in the restricted $M_X$
range yielded for the slope a value of $b^{MC(pN)} = 1.92 \pm 0.08$ in
good agreement with the value $b^{MCmeas} = 1.96 \pm 0.03$ obtained
before from a fit to the nondiffractive sample alone. The solid
straight lines show the nondiffractive contribution as obtained from
the fits; as before, they give a good description of the input
distributions for the nondiffractive contribution shown as the dashed
histograms and are in close agreement with the fits performed to the
nondiffractive contributions alone (dashed straight lines). Using the
fits and subtracting the nondiffractive contribution yielded for every
$(W,Q^2)$ interval the number of MC measured diffractive events; these
were found to be in very good agreement with the number of events in
the diffractive sample. This is a confirmation that the $\ln M^2_X$
method for extracting the diffractive contribution gives reliable
results also in the presence of diffractive contributions where the
proton dissociated.

Only a limited mass ($M_N$) range of $N^{dissoc}$ contributes to the
diffractive cross section presented below. This was studied with
EPSOFT. The distribution of the generated mass $M_N$ peaks at low
values, $M_N \laproe 2$ GeV. Due to the dominance of small $M_N$
values the secondaries from $N^{dissoc}$ are strongly collimated
around the direction of the incoming proton. Analysis of the angular
distribution of the secondaries as a function of $M_N$ showed that for
$M_N < 2$ GeV basically no energy is deposited in the calorimeter,
while for events with $M_N > 6$ GeV there are almost always
secondaries which deposit energy in the calorimeter. Furthermore,
events of the type $\gamma^* p \to XN^{dissoc}$, where decay particles
from $N^{dissoc}$ deposit energy in the calorimeter, have in general a
reconstructed mass which is substantially larger than the mass of $X$
and, therefore, these events contribute little to the event sample
selected for diffractive production of $\gamma^*p \to XN$.

As a result, for each $M_X$ bin there is an $M_N$ value, called
$M^{accept}_N$, such that the number of events with $M_N <
M^{accept}_N$, which migrate outside the bin is equal to those with
$M_N > M^{accept}_N$, which migrate into this bin. For the
($M_X,W,Q^2$) bins studied $M^{accept}_N$ was found to be 5.5 GeV to
within $\pm 1.5$ GeV. The spread in the value of $M^{accept}_N$
introduces an uncertainty in the cross section measurements which is
below the statistical uncertainty. Therefore, the diffractive cross
sections are quoted below as cross sections for $\gamma^*p \to XN$
with $M_N < 5.5$ GeV.

In the nominal analysis, the unfolding of the diffractive contribution
was performed considering only dissociation of the virtual photon as
simulated with RAPGAP. In order to test the sensitivity of the
unfolding to a contribution from double dissociation, the nominal
unfolding procedure was also applied to a MC sample consisting of a
mixture of single and double dissociation in the proportion given
above for the measured region. The resulting cross section gave a good
match to the cross section with $M_N < 5.5$ GeV as input.

\begin{figure}[ht]
\begin{center}
\epsfig{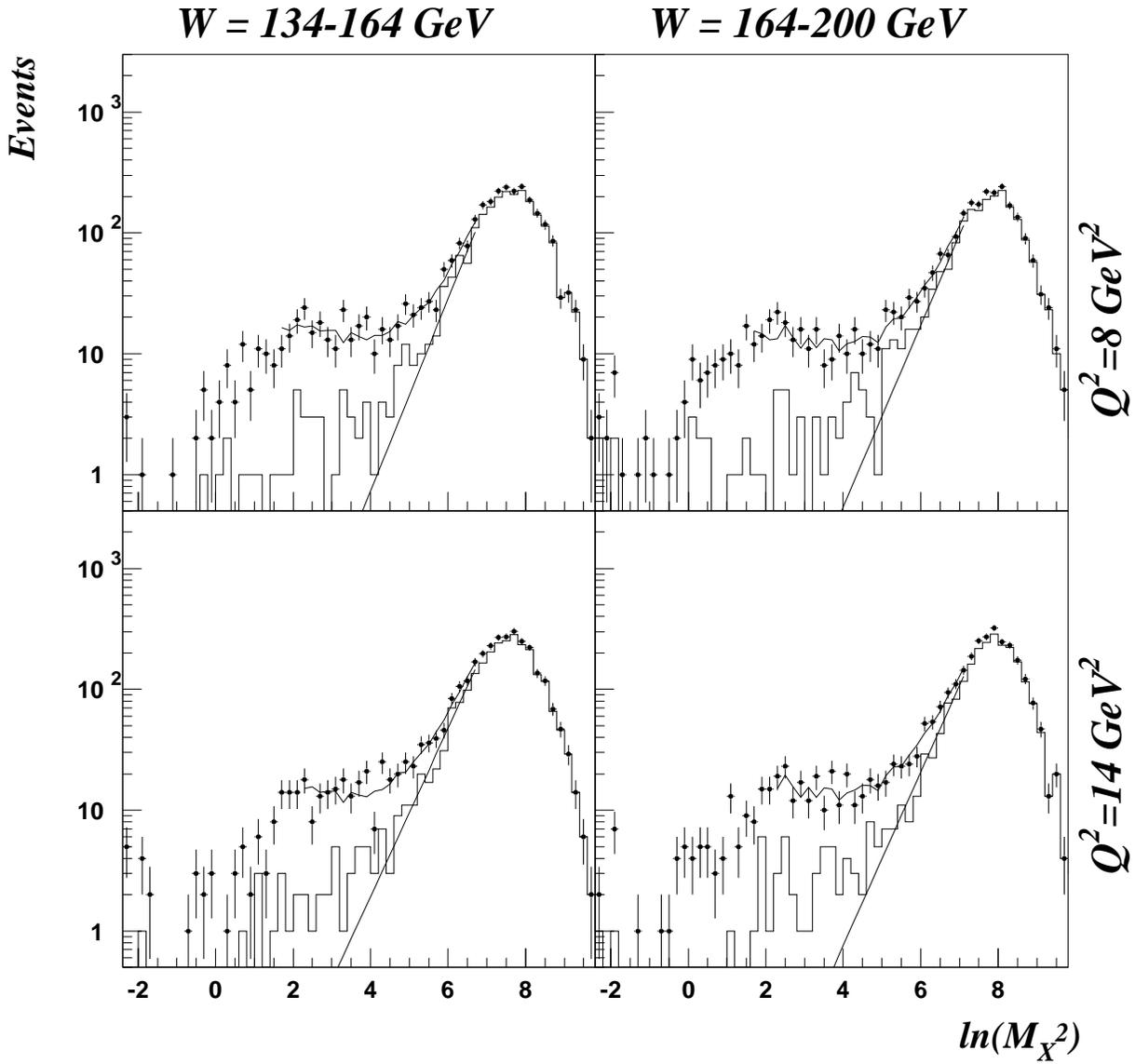}
\end{center}
\caption[]{Reaction $\gamma^* p \to X+anything$, where $X$ is the system
         observed in the detector. Distributions in $\ln M^2_X$ for
         data.  The straight lines give the nondiffractive
         contributions and the upper curves the sum of the diffractive
         and nondiffractive contributions as obtained from the fits.
         The solid histograms show the distributions for events with a
         PRT signal.}
\label{f:datafitl4prt}
\end{figure}

An attempt was made to estimate quantitatively the contribution from
proton dissociation by studying those events where a signal was
recorded in the PRT. The PRT registers particles emitted at large
pseudorapidities, $\eta = 4 - 6$. Such particles may result from
diffractive production where the proton dissociated or from
diffractive dissociation of the virtual photon into a system with a
large $M_X$, and from the proton remnant in nondiffractive
production. The points with error bars in Fig.~\ref{f:datafitl4prt}
show for the subset of the data with useful PRT information the $\ln
M^2_X$ distributions for all events. The straight lines show the
nondiffractive contribution as obtained from the fits performed in the
wide $M_X$ range. The sum of the diffractive and nondiffractive
contributions as obtained from the fits (upper curves) are in good
agreement with the data. The solid histograms in
Fig.~\ref{f:datafitl4prt} show the distributions (called PRT-tagged
distributions) for events with a PRT signal. Most of the events in the
high $\ln M^2_X$ peak have a PRT signal as expected for nondiffractive
scattering. The exponential fall-off of the PRT-tagged distributions
agrees well with the predictions obtained from the fits for the
nondiffractive contributions as shown by the straight lines. This
gives additional experimental proof for the reliability of the $M_X$
method.

A fraction of the events in the low mass region also has a PRT
signal. MC simulation showed that for $M_X < 3$ GeV almost all events
with a PRT signal are diffractive events where the proton
dissociated. From the number of PRT tagged events in this region,
using the efficiencies determined for the PRT counters~\cite{Beier}
and the prediction from EPSOFT for the PRT acceptance, events from
diffractive dissociation via $\gamma^* p \to XN^{dissoc}$ contribute
an estimated fraction of $XN^{dissoc}/XN = 27\pm 2 (stat)\%$ (plus an
unknown systematic uncertainty which depends on the mass spectrum and
decay properties of $N^{dissoc}$) to the diffractive cross sections
presented below. We prefer the estimate obtained by comparison with
the LPS analysis over that obtained here with the PRT since the former
does not rely on an assumption for the properties of $N^{dissoc}$.

\section{Evaluation of the cross sections}
For the final analysis, only bins where the fraction of nondiffractive
background was less than 50$\%$ and the purity was above 30$\%$ were
kept. Purity is defined as the ratio of the number of events generated
in the bin and observed in the same bin divided by the total number of
events observed in the bin. The average purity in the ($M_X,W,Q^2$)
bins was 52~$\%$.  The nondiffractive background fraction was small
for low $M_X$ or large $W$ values: for $M_X < 3$ GeV the background
fraction was typically below 2\% except for $Q^2 = 27$ (60) GeV$^2$
where in the lowest $W$ bin for which cross sections are presented it
amounted to 10\% (17\%). For $3 < M_X < 7.5$ GeV the background
fraction was typically below 5\% except for the next-to-lowest
(lowest) $W$ bin where it reached 18\% (33\%). For $7.5 < M_X < 15$
GeV the numbers were 22 - 37\% in the lowest bin and typically 15\%
elsewhere.

The average differential cross section for $ep$ scattering, in a given
($M_X, W, Q^2$) bin, was obtained by dividing the number of unfolded
events, ${\cal N}^{di\!f\!f}_{prod}$, by the luminosity, the bin
widths and the QED radiative correction factor. The lower limit of
$M_X$ was taken to be 2$m_{\pi}$, where $m_{\pi}$ is the pion mass.

The cross section for the process $ep \to e X N$ can be expressed in
terms of the transverse (T) and longitudinal (L) cross sections,
$\sigma_T^{di\!f\!f}$ and $\sigma_L^{di\!f\!f}$, for $\gamma^* p \to X
N$ as~\cite{Hand}:
\begin{eqnarray}
\frac{d\sigma^{di\!f\!f}_{\gamma^*p \to XN}(M_X,W,Q^2)}{dM_X}
\equiv \frac{d(\sigma_T^{di\!f\!f}+\sigma_L^{di\!f\!f})}{dM_X} \approx
\frac{2\pi}{\alpha}\,\frac{Q^2}{ (1-y)^2 +1}
 \,\frac{d\sigma^{di\!f\!f}_{ep \to eXN}(M_X,W,Q^2)}{dM_X d\ln W^2 dQ^2} \; .
\label{eq:sgp}
\end{eqnarray}
Here, a term $[1-\frac{y^2}{(1-y)^2+1}\frac{\sigma_L^{di\!f\!f}}
{\sigma_T^{di\!f\!f}+\sigma_L^{di\!f\!f}}]$ multiplying
[$\sigma_T^{di\!f\!f}+\sigma_L^{di\!f\!f}$] has been neglected. Since
$y\approx W^2/s$, this term can be substantially different from unity
only at high $W$ values. The effect is less than 5\% if $W < 158$ GeV
corresponding to $y \approx W^2 /s < 0.28$, or if $W < 200$ GeV and
$\sigma^{diff}_L < 0.5 \sigma^{diff}_T$.  In the extreme case that
$\sigma^{di\!f\!f}_L \gg
\sigma^{di\!f\!f}_T$, the term will increase $[\sigma^{di\!f\!f}_T +
\sigma^{di\!f\!f}_L ]$ by at most $11\%$ for the highest $W$ bin (164 - 200
GeV).

The differential cross section $d\sigma^{di\!f\!f}_{\gamma^* p \to
XN}/dM_X$ was determined from Eq.(\ref{eq:sgp}) in the different
($M_X$,$W$,$Q^2$) intervals using the fit results obtained assuming a
constant $D$ and transported to convenient values of $M_X$ and $Q^2$
using the shape of the parametrization of~\cite{Bartels98} (see
below).

{\bf Systematic errors}: The systematic uncertainties for the cross
section were estimated in a similar way as in the previous
work~\cite{Zepsdiff93} by varying the cuts and algorithms used to
select the events at the data and MC levels and repeating the full
analysis for every variant. The uncertainties in selecting the DIS
events arising from the identification of the scattered positron were
estimated by varying the box cut, the minimum energy and the minimum
probability required for the scattered positron. The uncertainties in
the reconstruction of the hadronic system were estimated by varying
the $y_{JB}$ cut between 0.01 and 0.03. In each case the deviations
from the nominal result were well below 10\% and typically 1 -
3\%. The uncertainty due to the beam-gas background was determined by
removing events without a reconstructed event vertex. This resulted in
small differences of up to 13\%. In order to test for remaining
background from photoproduction and the sensitivity to radiative
effects the cut on $\sum_j(E_j -p_{Zj})$ was varied. This resulted in
small changes except for three ($M_X,W,Q^2$) bins where differences of
up to 14\% were seen, commensurate with the corresponding statistical
errors.

Variation of the noise cuts for isolated calorimeter cells had a
negligible effect. Allowing for $\pm 5\%$ differences in the $M_X$
correction between data and MC simulation to account for uncertainties
in the energy scale of the calorimeter and the material in front
affected mainly the low $M_X$ bin, where at higher $Q^2$ values,
differences of up to 11\% were found. Note that increasing (lowering)
the mass correction factor systematically decreases (increases) the
cross section. Since the relative changes are basically independent of
$W$ this does not affect the $W$ dependence of the cross section. The
acceptance corrections and unfolding were also performed with a
different model for diffraction~\footnote{In this case diffraction was
modelled in RAPGAP using the parametrization developed
in~\cite{Zepsdiff93} on the basis of the model of~\cite{Nikzak}.}. For
$Q^2 < 27$ GeV$^2$ the differences were typically a few percent
reaching values between 11\% and 18\% in four bins. In the highest
$Q^2$ bin the differences were $\le 14\%$ except for the low $M_X$ bin
where they reached up to 57\%; this is mainly due to the small event
numbers involved.

The nominal fitting procedure for subtracting the nondiffractive
background used the fixed slope $b = b_{nom}$, see
Section~\ref{diffextraction}. The analysis was also performed with $b$
determined for the high $W$ bins 134~-~164, 164~-~200 GeV from the
wide $M_X$ range using for $D$ the extended form: $D = d_0 (1 - \beta)
[\beta (1 - \beta) + \frac{d_1}{2}(1 - \beta)^2]$, and the variable
form: $D = d_0 (1 - \beta) [\beta (1 - \beta) + \frac{d_1}{2}(1 -
\beta)^g]$,where $d_0, d_1$ and $g$ are parameters~\cite{Nikzak}
-~\cite{Bartels98}. Both forms consist of a quark-like (first term)
and a gluon-like (second term) contribution. In the wide $M_X$ range
the quark-like contribution decreases with $\ln M^2_X$ while the
gluon-like term increases. In the fits with the variable form the
gluon-like term dominates at large masses which leads to $D =
constant$ in the asymptotic limit. The resulting changes of the
diffractive cross section were small and did not exceed 13\%. For the
extended form, typical differences of 10 - 25\% were found; the
largest difference was found for one bin at $Q^2 = 60$ GeV$^2$ and
amounted to $-37\%$.

The total systematic error for each bin was determined by
quadratically adding the individual systematic uncertainties,
separately for the positive and negative contributions. The total
errors were obtained by adding the statistical and systematic errors
in quadrature.  The errors do not include an overall normalization
uncertainty of 2\% of which 1.5\% is from the luminosity determination
and 1.0\% from the uncertainty in the trigger efficiency.

\section{Differential cross section for $\gamma^*p \to X\;N$}

\subsection{$W$ and $Q^2$ dependence of $d\sigma^{di\!f\!f}_{\gamma^*p
 \to XN}/dM_X$ }
\begin{figure}[ht]
\begin{center}
\epsfig{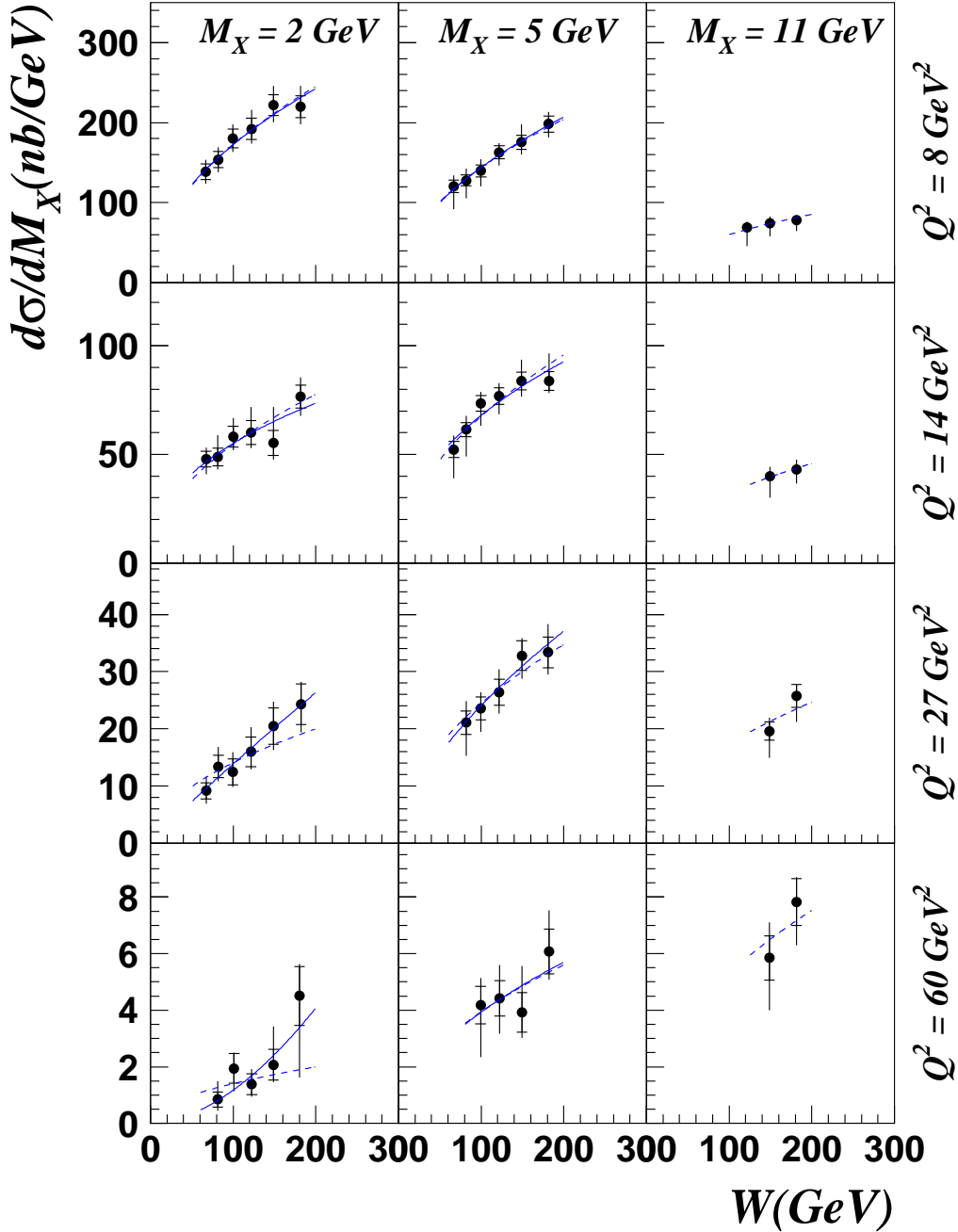}
\end{center}
\caption[]{The differential cross sections $d\sigma^{di\!f\!f}
          _{\gamma^* p \to X N}/dM_X, \; M_N < 5.5$ GeV, as a function
          of $W$ at average values of $M_X =$ 2, 5, 11~GeV, $Q^2 =$ 8,
          13, 27, 60~GeV$^2$. The inner error bars show the
          statistical errors and the full bars the statistical and
          systematic errors added in quadrature.  The overall
          normalization uncertainty of 2\% is not included.  The solid
          curves show the result from fitting the diffractive cross
          section for each ($W,Q^2$) bin separately using the form
          $d\sigma^{di\!f\!f}_{\gamma^* p \to X N}/dM_X \propto
          (W^2)^{a^{diff}}$ where $a^{diff}$ and the normalization
          constants were treated as free parameters. The dashed curves
          show the result from the fit where $a^{diff}$ was assumed to
          be the same for all ($W,Q^2$) bins.}
\label{f:sigwfinal}
\end{figure}

\begin{figure}[ht]
\begin{center}
\epsfig{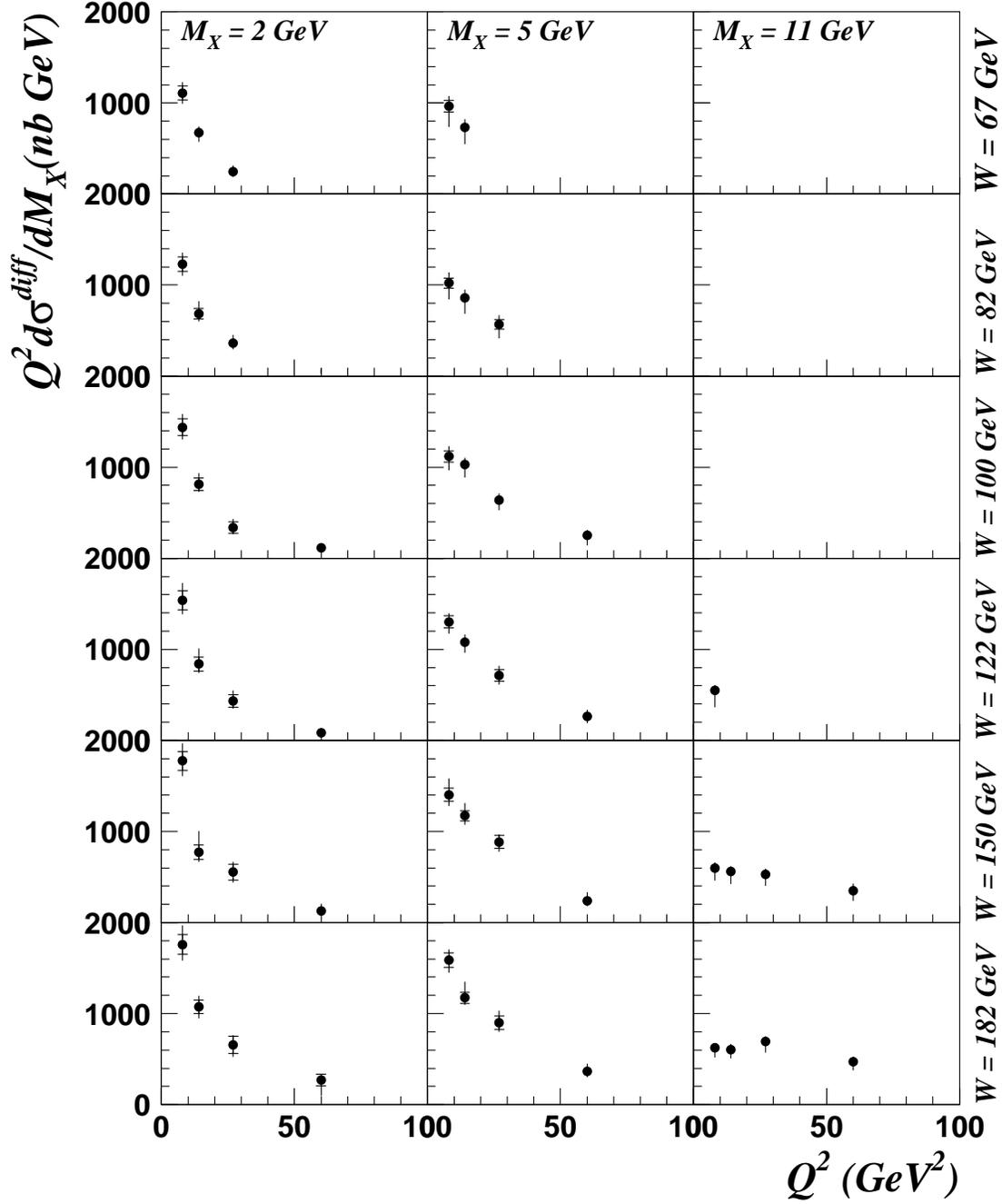}
\end{center}
\vspace{-0.6cm}
\caption[]{The diffractive differential cross section
         for $\gamma^* p \to XN, M_N < 5.5$ GeV, multiplied by $Q^2$,
         $Q^2 d\sigma^{di\!f\!f}(\gamma^* p \to X N)/dM_X$, as a
         function of $Q^2$ for the values of $M_X$ and $W$ indicated.
         The inner error bars show the statistical errors and the full
         bars the statistical and systematic errors added in
         quadrature.}
\label{f:sigqfinal}
\end{figure}

The diffractive cross section for $\gamma^* p \to XN, \:(M_N < 5.5$
GeV), is presented in Table~\ref{t:diffcrosstab} and
Fig.~\ref{f:sigwfinal} as a function of $W$ for various $M_X$ and
$Q^2$ values. From comparison with published data~\cite{Zeprho} -
~\cite{H1epv}, about 20\% of the diffractive cross section observed in
the mass bin $M_X < 3$~GeV at $7 < Q^2 < 20$ GeV$^2$ results from the
production of the vector mesons $V= \rho^0, \phi$ via $\gamma^* p \to
V N$. In Fig.~\ref{f:sigqfinal} $Q^2 d\sigma^{di\!f\!f}_{\gamma^*p \to
XN}/dM_X $ is presented as a function of $Q^2$ for different ($M_X,W$)
values. A fast decrease with $Q^2$ is observed for small $M_X$ which
is similar to the behaviour of DIS vector meson
production~\cite{Zeprho} - ~\cite{H1epv}. The decrease becomes slower
for the high $M_X$ region showing that high mass dissociation of the
virtual photon becomes increasingly more important as $Q^2$ grows. For
a discussion in terms of a partonic behaviour see below.

The diffractive cross section rises rapidly with $W$ at all $Q^2$
values for the $M_X$ bins up to 7.5~GeV. The cross section was
fitted~\cite{Blobel} for each ($M_X,Q^2$) bin using the form
\begin{eqnarray}
\frac {d\sigma^{di\!f\!f}_{\gamma^*p \to XN}(M_X,W,Q^2)}{ dM_X } & =
h \cdot W^{a^{diff}} \; \; \; ,
\label{eq:gxpint}
\end{eqnarray}
where $a^{diff}$ and the normalization constants $h$ were treated as
free parameters. The form given in Eq.~\ref{eq:gxpint} with a single
value of $a^{diff}$ gives an acceptable $\chi^2$ value ($\chi^2
/d.o.f. = 37/35 = 1.1$), considering only the statistical errors)
although one observes in Fig.~\ref{f:sigwfinal} (see dashed curves) a
tendency for the data at low $M_X$, high $Q^2$ to have a steeper slope
than resulting from this fit. The fit yielded $a^{diff} = 0.507 \pm
0.034 (stat)^{+0.155}_{-0.046} (syst)$. Here, the largest contribution
to the positive systematic error arises when the extended form is used
for $D$ in the determination of the slope $b$. In Regge models,
$a^{diff}$ is related to the trajectory of the pomeron $\alphapom
(t)$, averaged over $t$: $\overline{\alphapom} = 1+a^{diff}/4$. The
fit value for $a^{diff}$ leads to $\overline{\alphapom} = 1.127 \pm
0.009 (stat)^{+0.039}_{-0.012}(syst)$.

\begin{figure}[ht]
\begin{center}
\epsfig{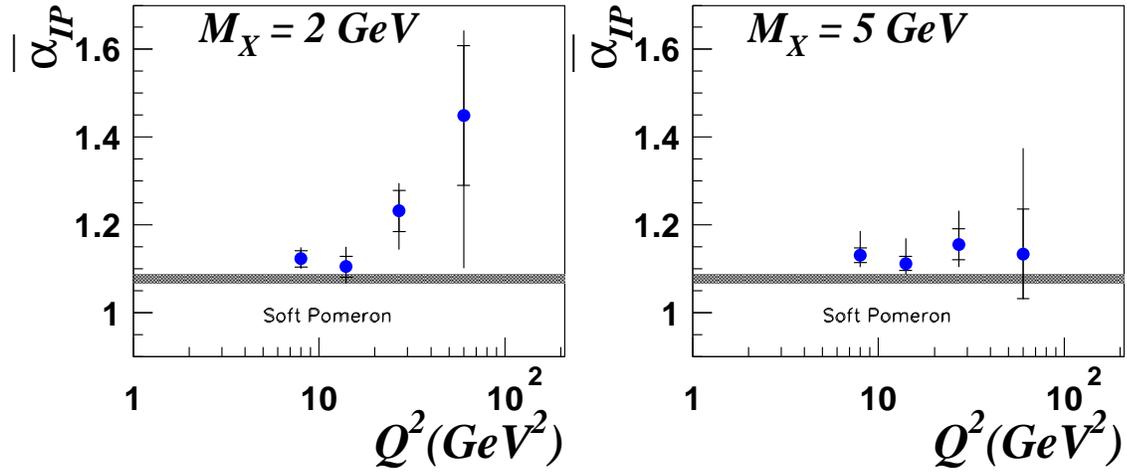}
\end{center}
\vspace{3cm}
\caption[]{The parameter $\overline{\alphapom}$ obtained from the fits
         versus $Q^2$ for $M_X = 2$ and 5 GeV.  The inner error bars
         show the statistical errors and the full bars the statistical
         and systematic errors added in quadrature.  The result for a
         soft pomeron is indicated by the heavy horizontal band.}
\label{f:apfinal}
\end{figure}

In order to test for a possible $Q^2$ dependence of
$\overline{\alphapom}$ a fit was performed where $a^{diff}$ was taken
as a free parameter for every ($M_X, Q^2$) bin. This resulted also in
a good description of the data (see solid curves in
Fig.~\ref{f:sigwfinal}). The resulting $\overline{\alphapom}$ values
are shown in Fig.~\ref{f:apfinal} with the statistical and systematic
uncertainties added in quadrature. The systematic uncertainties were
estimated by repeating the fits independently for every source of
systematic error.  With the present accuracy of the data, no
conclusion can be drawn on a possible $Q^2$ dependence of
$\overline{\alphapom}$.

In deriving the diffractive cross section, the assumption was made
that the term\newline
$[1-\frac{y^2}{(1-y)^2+1}\frac{\sigma_L^{di\!f\!f}}
{\sigma_T^{di\!f\!f}+\sigma_L^{di\!f\!f}}]$ can be neglected (see
Eq.~\ref{eq:sgp}). This holds when $\sigma^{diff}_L \ll
\sigma^{diff}_T$. If the assumption is made that $\sigma^{diff}_L =
\sigma^{diff}_T$ ($\sigma^{diff}_T \ll \sigma^{diff}_L$) the value of
$a^{diff}$ increases by 0.047 (0.096) and hence $\overline{\alphapom}$
increases by 0.012 (0.024).

H1~\cite{H1epf2d394} has given for the intercept of the pomeron
trajectory a value of $\alphapom(0) = 1.203 \pm 0.020(stat)\pm
0.013(syst)$$^{+0.030}_{-0.035}(model)$. Averaging over the
$t$-distribution~\footnote{When parametrizing the $t$-distribution by
$d\sigma /dt \propto exp[(b_0+2\alphapom^{\prime} \ln(1/x_{\pom}))t]$
the difference $\alphapom (0) - \overline{\alphapom}$ is determined
mainly by $b_0$ and $\alphapom^{\prime}$. H1 quotes $b_0=4.6$
GeV$^{-2}$ and $\alphapom^{\prime} = 0.26$ GeV$^{-2}$.} gives
approximately $\overline{\alphapom} = \alphapom (0) - 0.03$, a value
which is consistent with the result from this analysis~\footnote{In
the analysis of H1~\cite{H1epf2d394} the diffractive contribution was
first extracted by selecting events with a large rapidity gap. From
the $\xpom$ dependence of the resulting diffractive structure function
$F^{D(3)}_2(\xpom,\beta,Q^2)$ it was concluded that at large $\xpom$ a
substantial reggeon contribution was present. The latter was
determined from a Regge fit to the data using the sum of a pomeron and
a reggeon contribution. We have performed a similar fit to our data
and found no evidence for a significant contribution from reggeon
exchange.}.

Our value of $\overline{\alphapom} = 1.127 \pm 0.009
(stat)^{+0.039}_{-0.012}(syst)$ lies above the results deduced from
hadron-hadron scattering where the intercept of the pomeron trajectory
was found to be $\alphapom^{soft}~(0)~= 1.08$~\cite{Donlan1} and
$\alphapom^{soft}~(0)~=~1.096^{+0.012}_{-0.009}$~\cite{Cudell}. Averaging
over $t$ reduces these values by about 0.02~\footnote{Assuming for the
$t$-distribution $d\sigma /dt \propto exp[(b_0+2\alphapom^{\prime} \ln
W^2) t]$ with $b_0 = 7$ GeV$^{-2}$ (from $\pi^-p$ elastic scattering,
see~\cite{Goulianos}) and $\alphapom^{\prime} = 0.25$ GeV$^{-2}$
~\cite{Donlan1}.}leading to $\overline{\alphapom}^{soft} =
1.06$~\cite{Donlan1} and
$\overline{\alphapom}^{soft}~=~1.076^{+0.012}_{-0.009}$,
respectively. The latter value is shown by the horizontal bands in
Fig.~\ref{f:apfinal}.

\subsection{Comparison of the diffractive and total cross sections}
\begin{figure}[ht]
\begin{center}
\epsfig{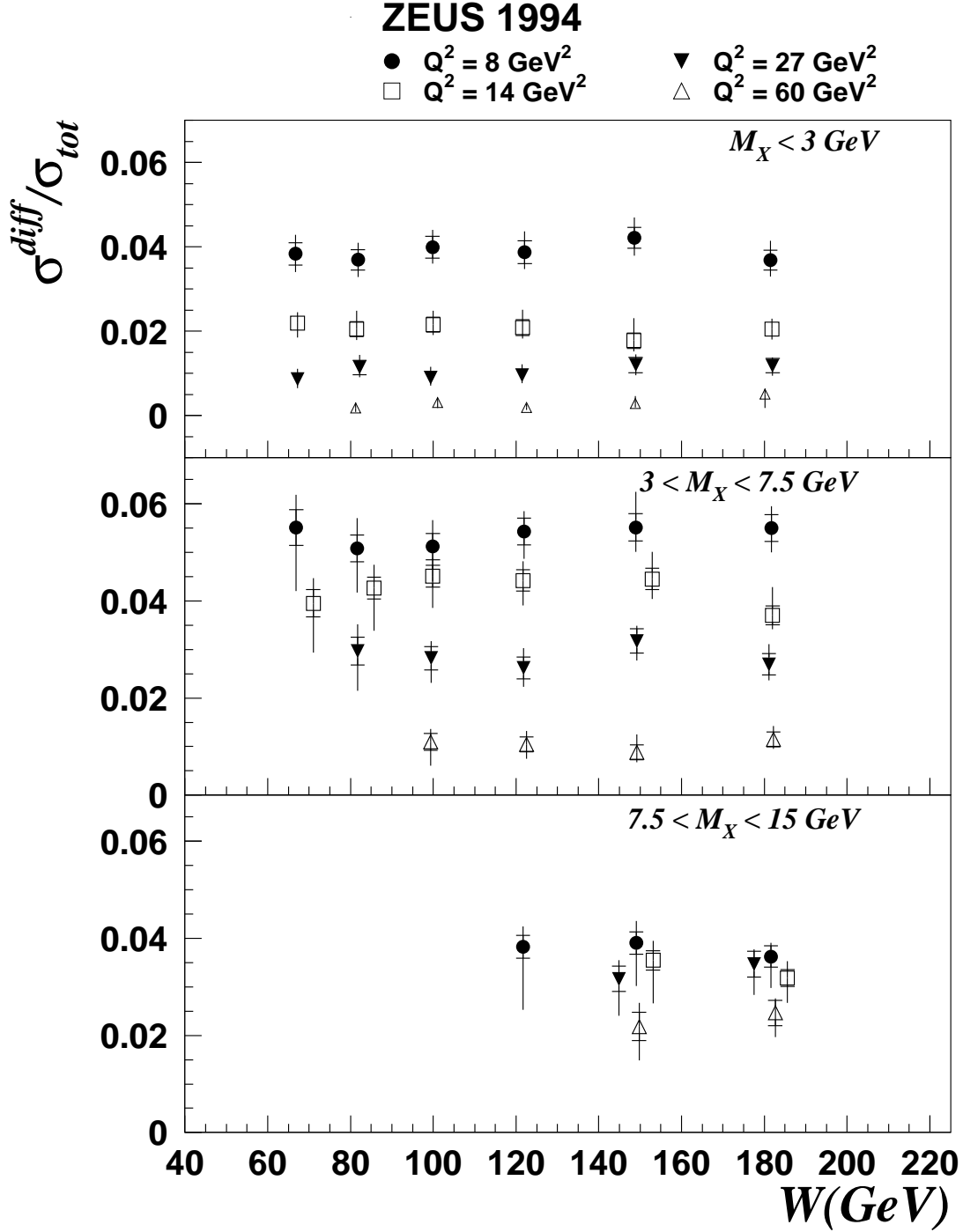}
\end{center}
\caption[]{The ratio of the diffractive cross section,
         integrated over the $M_X$ intervals indicated, $\sigma^{diff}
         = \int^{M_b}_{M_a} dM_X\sigma^{diff}_{\gamma^* p\to XN}$, for
         $M_N < 5.5$ GeV, to the total cross section for virtual
         photon proton scattering, $r^{diff}_{tot} =
         \sigma^{diff}/\sigma_{\gamma^* p}^{tot}$, as a function of
         $W$ for the $M_X$ intervals and $Q^2$ values indicated.
         $\sigma^{\gamma^* p}_{tot}$ was taken from our $F_2$
         measurements using the 1994 data \protect\cite{Zepf294}.  The
         inner error bars show the statistical errors and the full
         bars the statistical and systematic errors added in
         quadrature.  }
\label{f:sigwfinalstot}
\end{figure}

The ratio of the diffractive cross section to the total virtual-photon
proton cross section,
\begin{eqnarray}
r^{diff}_{tot}\;=\; \frac{{\bf\int^{M_b}_{M_a}} dM_X
d\sigma^{di\!f\!f}_{\gamma^* p \to XN}/dM_X}{\sigma_{\gamma^* p}^{tot} },
\end{eqnarray}
is displayed in Fig.~\ref{f:sigwfinalstot} as a function of $W$ for
the different $M_X$ bins ($M_a < M_X < M_b$) and $Q^2$ values. The
total cross section was taken from our $F_2$ measurements performed
with the 1994 data~\cite{Zepf294} using $\sigma_{\gamma^*
p}^{tot}(W,Q^2) = \frac{4 \pi^2 \alpha}{Q^2(1-x)} F_2(x \approx
\frac{Q^2}{W^2},Q^2)$. The data show that, for fixed $M_X$, the
diffractive cross section possesses the same $W$ dependence as the
total cross section. A fit of $r^{diff}_{tot}$ using the form
$r^{diff}_{tot} \propto W^{\rho}$, considering all data with $M_X <
7.5$ GeV and treating $\rho$ and the normalization constants for every
($M_X,Q^2$) interval as free parameters, yielded $\rho = 0.00 \pm
0.03(stat)$, consistent with $W$ independence. The same conclusion is
reached when comparing the value of the power $a^{diff} = 0.507 \pm
0.034(stat) ^{+0.155}_{-0.046}(syst)$ with the power $a^{tot} = 0.55
\pm 0.02$ obtained for $\sigma_{\gamma^* p}^{tot}$ in the same
($W,Q^2$) range. Equal powers for the diffractive and total cross
sections is contrary to the na\"{\i}ve expectation. Assuming (i) for
DIS the validity of the optical theorem~\cite{Brodsky97} and (ii) that
the cross section for diffractive photon dissociation at fixed $M_X$
has the same $W$ dependence as the forward cross section for elastic
scattering, $\gamma^* p \to \gamma^*p$, then
$d\sigma^{di\!f\!f}_{\gamma^* p \to XN}/dM_X$ should be proportional
to $W^a$ if $\sigma_{\gamma^* p}^{tot} \propto W^{a/2}$. Hence, taking
the $W$ dependence found for $d\sigma^{di\!f\!f}_{\gamma^* p \to
XN}/dM_X$ the power $\rho$ would have been expected to be $\rho =
a^{diff}/2 = 0.25 \pm 0.02(stat)^{+0.07}_{-0.02}(syst)$, in clear
disagreement with the data.

The rapid rise of $\sigma_{tot}$ with $W$, which is equivalent to the
rapid rise of $F_2$ as $x \to 0$, in QCD is attributed to the
evolution of partonic processes. The observation of similar $W$
dependences for the total and diffractive cross sections suggests,
therefore, that diffraction in DIS receives sizeable contributions
from hard processes. The same $W$ dependence for the diffractive and
total cross sections was predicted in~\cite{Buchmuller95} (see below).

The diffractive contribution to the total cross section for low $M_X$
decreases rapidly as $Q^2$ increases while for $M_X = 11$ GeV it is
the same, within a factor of two. Averaged over $W = 60 - 200$ GeV
(considering only the accepted bins), the diffractive contribution
(with $M_N < 5.5$ GeV) to the total cross section for $M_X < 7.5$ GeV
is $9.2^{+0.3}_{-0.4}\%$ ($6.3^{+0.3}_{-0.3}\%, 3.9^{+0.2}_{-0.3}\%,
1.3^{+0.2}_{-0.2}\%$) at $Q^2 = 8 (14, 27, 60)$ GeV$^2$. The
corresponding fractions for $M_X < 15$ GeV in the $W$ range 134 - 200
GeV are $13.2^{+0.5}_{-0.7}\%$ ($9.4^{+0.5}_{-0.6}\%,
7.5^{+0.3}_{-0.5}\%, 3.7^{+0.3}_{-0.4}\%$). As $Q^2$ increases, the
high $M_X$ region becomes more important.

 The $Q^2$ behaviour of the diffractive contribution to the total
 cross section observed here for $M_X < 15$ GeV is not in
 contradiction to our earlier finding that the fraction of DIS events
 with a large rapidity gap (LRG) is rather constant with
 $Q^2$~\cite{Zeplrg93}: the LRG analysis considered events with masses
 $M_X \gaproe 4$ GeV, did not impose an upper limit on $M_X$ and
 included the $W$ range up to 270 GeV.

\section{Diffractive structure function of the proton}
The concept of a diffractive structure function introduced
in~\cite{Ingelmanschlein} is based on the assumption that diffraction
is mediated by the exchange of a colourless object, called a pomeron,
which is composed of partons.  The diffractive structure function of
the proton can be related to the diffractive cross section as
follows~\cite{Ingelman}:
\begin{eqnarray}
\frac{1}{2M_X}\frac{d\sigma^{di\!f\!f}_{\gamma^*p \to XN}(M_X,W,Q^2)}{dM_X} =
4\pi^2\alpha \frac{W^2}{(Q^2 +W^2)^2 Q^2} F^{D(3)}_2(\beta,\xpom,Q^2).
\label{eq:f2d3sdiff}
\end{eqnarray}
For $W^2 \gg Q^2$, which holds for this analysis,
Eq.~\ref{eq:f2d3sdiff} can be written as:
\begin{eqnarray}
\frac{1}{2M_X}\frac{d\sigma^{di\!f\!f}_{\gamma^*p \to XN}(M_X,W,Q^2)}{dM_X} \approx
  \frac{4\pi^2\alpha}{Q^2(Q^2 +M^2_X)} \xpom F^{D(3)}_2(\beta,\xpom,Q^2).
\label{eq:f2d3sdiffapprox}
\end{eqnarray}
If $F^{D(3)}_2$ is interpreted in terms of quark densities then it
specifies for a diffractive process the probability to find a quark
carrying a momentum fraction $x = \beta \xpom$ of the proton momentum.

It has been suggested~\cite{Ingelmanschlein} that
$F^{D(3)}_2(\xpom,\beta,Q^2)$ should factorize into a term which
depends on the probability of finding a pomeron carrying a fraction
$\xpom$ of the proton momentum and the pomeron structure function
$F^{D(2)}_2$ given in terms of the pomeron's quark densities which
depend on $\beta$ and $Q^2$:
\begin{eqnarray}
F^{D(3)}_2(\xpom,\beta,Q^2) = f_{\pom}(\xpom)F^{D(2)}_2(\beta,Q^2)
\end{eqnarray}
where $f_{\pom}(\xpom)$ is generically called the pomeron flux factor.

The quantity $\xpom F^{D(3)}_2$ is given in Table~\ref{t:xf2d3tab} and
shown in Fig.~\ref{f:xf2dfinaltheo} as a function of $\xpom$ for
different combinations of $\beta$ and $Q^2$.

\begin{figure}[ht]
\begin{center}
\epsfig{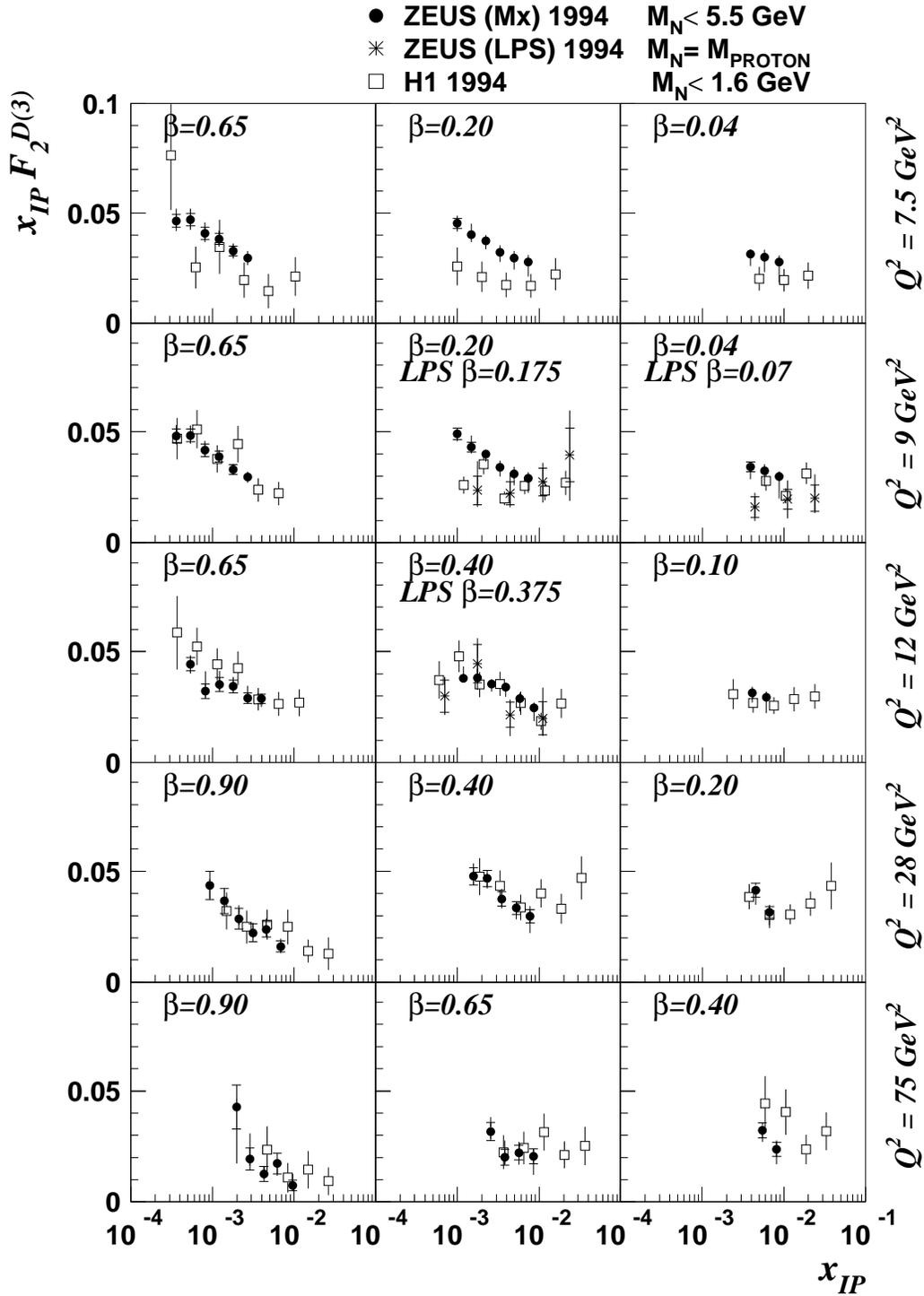}
\end{center}
\vspace{-0.6cm}
\caption[]{The diffractive structure function of the proton for 
         $\gamma^* p \to XN, M_N < 5.5$ GeV, multiplied by $\xpom$,
         $\xpom F^{D(3)}_2(\xpom,\beta,Q^2)$, from this analysis
         (solid points) compared with the results from our previous
         LPS measurement obtained with an identified proton for
         $\gamma^* p \to Xp$ (stars) and from a subsample of the H1
         data (open points) for $\gamma^* p \to XN, M_N < 1.6$
         GeV. For ease of comparison the results from this analysis
         are scaled to ($\beta, Q^2$) values used by H1.  The data
         points from this experiment shown for $Q^2 = 7.5$ and 9
         GeV$^2$ are those obtained at $Q^2 = 8$ GeV$^2$ shifted to
         $Q^2 = 7.5$ and 9 GeV$^2$. The LPS data are given for
         slightly different $\beta$ and $Q^2$ values.  }
\label{f:xf2dfinalh1lps}
\end{figure}

In Fig.~\ref{f:xf2dfinalh1lps} the data from this analysis (solid
points) are compared with ZEUS data obtained using the Leading Proton
Spectrometer (LPS)~\cite{Zepf2d3lps94} and with those of
H1~\cite{H1epf2d394}. For ease of comparison the $\xpom F^{D(3)}_2$
values from this analysis were scaled to the ($\beta,Q^2$) values used
in the H1 analysis. The LPS data correspond to events of the type
$\gamma^*p \to Xp$ with an identified proton. No correction was
applied for the contribution from double dissociation which is present
in this analysis but not in the LPS data. The correction would
increase the LPS data by a factor of $1.45^{+0.34}_{-0.23}$. There is
consistency between this analysis and the LPS data. The H1 data
correspond to $M_N < 1.6$ GeV while those from this analysis are given
for $M_N < 5.5$ GeV. No correction was applied. The data from H1
approximately agree with those from this analysis. However, for fixed
$\beta$, the H1 values have a tendency to rise faster with $Q^2$ even
allowing for an extra scaling factor.

\subsection{$\xpom$ dependence of $\xpom F^{D(3)}_2(\xpom,\beta,Q^2)$}

 The data from this analysis (Fig.~\ref{f:xf2dfinaltheo}) show that
 $\xpom F^{D(3)}_2(\xpom,\beta,Q^2)$ decreases with increasing
 $\xpom$, which reflects the rapid increase of the diffractive cross
 section with rising $W$. Assuming the flux factor to be of the form
 $f_{\pom}(\xpom) = (C/\xpom)\cdot (x_0/\xpom)^n$, taking for the
 arbitrary normalization constant $C = 1$ and for $x_0$ the average
 value of the measured $\xpom$, $x_0 = 0.0042$, the data were fitted
 with the form $\xpom F^{D(3)}_2(\xpom,\beta,Q^2) =(C/\xpom)\cdot
 (x_0/\xpom)^n F^{D(2)}_2(\beta,Q^2)$, which leads to
 $F^{D(2)}_2(\beta,Q^2) = x_0 F^{D(3)}_2(x_0,\beta,Q^2)$. The values
 $F^{D(2)}_2(\beta_i,Q^2_i)$ at the 12 measured ($\beta_i,Q^2_i$)
 points and $n$ were treated as fit parameters. A good fit was
 obtained ($\chi^2/d.o.f. = 41/40 = 1.0$, statistical errors only)
 yielding $n = 0.253 \pm 0.017(stat)^{+0.077}_{-0.023}(syst)$.  Note
 that $n \simeq a^{diff}/2 = 2(\overline{\alphapom}-1)$. The fit was
 also performed assuming $n$ to depend logarithmically on $Q^2$. This
 resulted in small differences that were included in the errors given
 for $F^{D(2)}_2$. The fact that a good fit was found with a single
 value for $n$ shows that the data are consistent with the assumption
 that $F^{D(3)}_2$ factorizes into a flux factor depending only on
 $\xpom$ and a structure function $F^{D(2)}_2$ which depends on
 ($\beta,Q^2$). Note that there is an arbitrary normalization factor
 for the flux and therefore also for $F^{D(2)}_2$.

\subsection{$\beta$ and $Q^2$ dependence of $\xpom F^{D(3)}_2(\xpom,\beta,Q^2)$
and $F^{D(2)}_2(\beta,Q^2)$}

\begin{figure}[ht]
\begin{center}
\epsfig{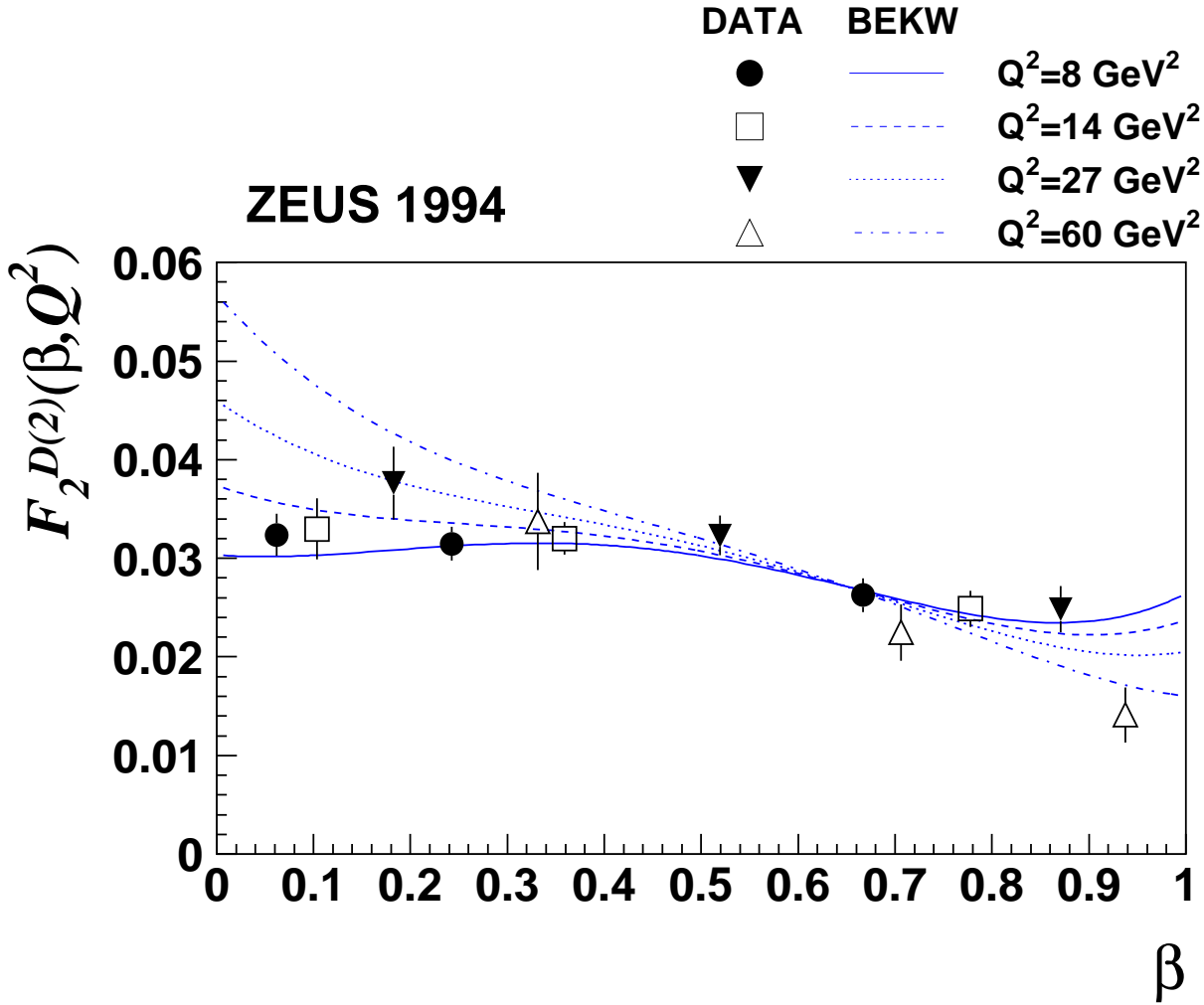}
\end{center}
\vspace{-0.6cm}
\caption[]{The structure function $F^{D(2)}_2(\beta,Q^2)$
         for $\gamma^* p \to XN, M_N < 5.5$ GeV, for the $Q^2$ values
         indicated, as a function of $\beta$ as extracted from a fit
         to the measured $\xpom F^{D(3)}_2$ values, see text.  The
         error bars show the statistical and systematic errors added
         in quadrature.  The curves show the fit results obtained with
         the BEKW model discussed in the text.}
\label{f:f2dvsbeta}
\end{figure}

The $F^{D(2)}_2(\beta,Q^2)$ values obtained from the fit described
above are presented in Fig.~\ref{f:f2dvsbeta} as a function of $\beta$
for all $Q^2$ values. It should be stressed that these $F^{D(2)}_2$
values do not depend on whether the $F^{D(3)}_2$ factorizes into a
pomeron flux factor or not since a) $F^{D(2)}_2(\beta,Q^2) = x_0
F^{D(3)}_2(x_0,\beta,Q^2)$ and the fit was basically only used to
interpolate to $\xpom = x_0$; b) a fit with a $Q^2$ dependent flux
gave basically the same $F^{D(2)}_2$ values. The data show that the
diffractive structure function $F^{D(2)}_2$ has a simple
behaviour. For $\beta < 0.6$ and $Q^2 < 14$ GeV$^2$, $F^{D(2)}_2$ is
approximately independent of $\beta$.  For $\beta < 0.8$ also the data
from different $Q^2$ values are rather similar suggesting a leading
twist behaviour characterized by a slow $\ln Q^2$ type rescaling. For
$\beta > 0.9$ the data show a decrease with $\beta$ or $Q^2$.

The $Q^2$ behaviour of $\xpom F^{D(3)}_2(\xpom,\beta,Q^2)$ is shown as
solid points in Fig.~\ref{f:xf2d3f2vsq2}. The data are presented for
fixed ($M_X,W$), the variables in which the diffractive contribution
was extracted. Given ($M_X,W$) and $Q^2$ the value of $\xpom$ can be
calculated. For $M_X < 7.5$ GeV, $\xpom F^{D(3)}_2$ decreases with
$Q^2$ while for $M_X = 11$ GeV it is approximately constant.

Strong $Q^2$ variations, which are found e.g. for the diffractive
cross section (see Fig.~\ref{f:sigqfinal}), are just a reflection of
kinematics: the strong $Q^2$ variation of
$Q^2d\sigma^{di\!f\!f}_{\gamma^*p \to XN}/dM_X$ is mainly controlled
by the kinematical factor $M_X/(Q^2 + M^2_X)$ in
Eq.~\ref{eq:f2d3sdiffapprox}.

The approximate constancy of $F^{D(2)}_2$ for $\beta < 0.9$ combined
with the rapid rise of $F^{D(3)}_2$ as $\xpom$ decreases can be
interpreted as evidence for a substantial partonic component in DIS
diffraction dissociation.

\subsection{Comparison with models}

The diffractive process in DIS has attracted considerable attention
because of the possibility that this process can be described by
perturbative QCD (pQCD). In parton models the process can be
visualized as a fluctuation of the incoming virtual photon into a
$q\overline{q}$ pair followed by the interaction of this pair with the
incoming proton leading to a $q\overline{q}$ state plus, well
separated in rapidity, a proton or debris from the dissociation of the
proton. In~\cite{Bjorken} it was argued that the dominant contribution
to diffraction in DIS comes from the aligned jet configuration where
$q$ and $\overline{q}$ from photon dissociation have small transverse
momenta relative to the direction of the virtual photon leading to the
same energy dependence as observed for diffraction in hadron-hadron
scattering. This contribution was expected to scale with $Q^2$. The
$\beta$ distribution for the aligned jet configuration from transverse
photons was predicted~\cite{Donnachie87} to be of the form
\begin{eqnarray}
F^T_{q\overline{q}}\propto \beta(1-\beta).
\label{qqterm}
\end{eqnarray}

The same $\beta$ dependence was expected in pQCD when the aligned
quarks interact with the proton through two-gluon
exchange~\cite{Nikzak}. The production of a $q\overline{q}g$ system by
transverse photons was also found to be leading twist and was assumed
to have a $\beta$ dependence of the type~\cite{Nikzak}
\begin{eqnarray}
F^T_{q\overline{q}g} \propto (1-\beta)^{\gamma}
\label{qqgterm}
\end{eqnarray}
with $\gamma = 2$. A later calculation~\cite{Wusthoff} found $\gamma =
3$. In the same approach the contribution to the production of a
$q\overline{q}$ system by longitudinal photons was found to be of
higher twist and to have a $\beta$ dependence of the form
\begin{eqnarray}
F^L_{q\overline{q}} \propto \beta^3(1-2\beta)^2.
\label{longterm}
\end{eqnarray}
In pQCD models the $\xpom$ dependence is expected to be driven by the
$x$ dependence of the square of the gluon momentum density of the
proton~\cite{Ryskin}, $[x\cdot g(x,\mu^2)]^2$, with $x=\xpom$ and
$\mu$ is the probing scale.

In~\cite{Gotsman97} the sum of the contributions from the three terms
$F^T_{q\overline{q}}, F^L_{q\overline{q}}, F^T_{q\overline{q}g}$ was
evaluated in the perturbative region.

\begin{figure}[ht]
\begin{center}
\epsfig{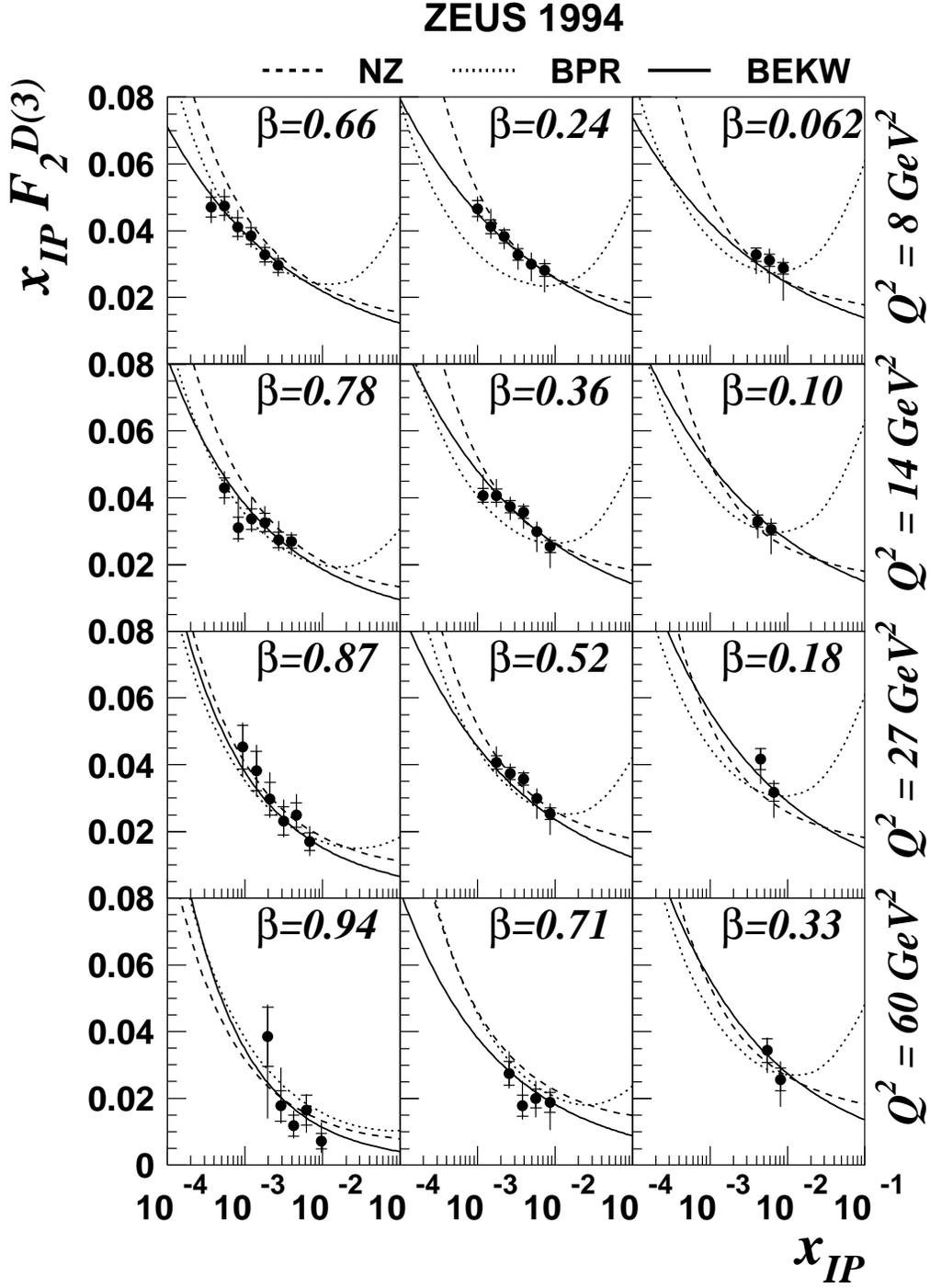}
\end{center}
\vspace{-0.6cm}
\caption[]{The diffractive structure function of the proton multiplied
         by $\xpom$, $\xpom F^{D(3)}_2$, as a function of $\xpom$ from
         this analysis (solid points).  The inner error bars show the
         statistical errors and the full bars the statistical and
         systematic errors added in quadrature.  The curves show the
         results from the models of Nikolaev and Zhakarov (NZ),
         Bialas, Peschanski and Royon (BPR) and Bartels, Ellis,
         Kowalski and W\"{u}sthoff (BEKW).  }
\label{f:xf2dfinaltheo}
\end{figure}

We now compare the data with three partonic models (NZ)~\cite{Nikzak},
(BPR)~\cite{Bialas} and (BEKW)~\cite{Bartels98}. In the NZ model,
diffractive dissociation is described as a fluctuation of the photon
into a $q\overline{q}$ or $q\overline{q}g$ Fock
state~\cite{Nikzak}. The interaction with the proton proceeds via the
exchange of a BFKL~\cite{BFKL} type pomeron, starting in lowest order
from the exchange of a two-gluon system in a colour-singlet state. The
BPR model describes the process $\gamma^* p \to Xp$ as the scattering
of a colour dipole from the photon on a colour dipole from the
proton. The model parameters were chosen by comparison with the H1
data~\cite{H1epf2d394}. The predictions of the NZ and BPR models are
shown in Fig.~\ref{f:xf2dfinaltheo} for $\xpom F^{D(3)}_2$ as a
function of $\xpom$ by the dashed and dotted curves, respectively. The
NZ model provides a reasonable description of the data. The BPR model
has some difficulties in reproducing the data for medium values of
$\beta$ and $Q^2 < 14$ GeV$^2$.

We studied the individual contributions from the three terms in
Eqs.~\ref{qqterm}-~\ref{longterm} following~\cite{Bartels98} which
identified them as the major contributors to the diffractive structure
function. In~\cite{Bartels98} they were calculated in the perturbative
region and extended into the soft region. The $\xpom$ dependence was
assumed to be of the form $(1/\xpom)^n$. The power $n$ was allowed to
be different for the transverse ($n_T$) and the longitudinal ($n_L$)
contributions. The normalizations of the three terms were determined
from the data. This is called the BEKW model in the following:
\begin{eqnarray}
\xpom F^{D(3)}_2(\beta,\xpom,Q^2) = c_T \cdot F^T_{q\overline{q}} +
c_L \cdot F^L_{q\overline{q}} + c_g \cdot F^T_{q\overline{q}g}
\end{eqnarray}
with
\begin{eqnarray}
F^T_{q\overline{q}} & = & (\frac{x_0}{\xpom})^{n_T(Q^2)} \cdot \beta (1-\beta) \\
F^L_{q\overline{q}} & = & (\frac{x_0}{\xpom})^{n_L(Q^2)} \cdot \frac{Q^2_0}{Q^2}
\cdot [\ln (\frac{7}{4}+\frac{Q^2}{4\beta Q^2_0})]^2 \cdot \beta^3(1-2\beta)^2 \\
F^T_{q\overline{q}g} & = & (\frac{x_0}{\xpom})^{n_T(Q^2)} \cdot
\ln (1+\frac{Q^2}{Q^2_0}) \cdot (1-\beta)^{\gamma} \\
n_{T,L}(Q^2) & = & 0.1 +  n^0_{T,L} \cdot \ln [1+\ln(\frac{Q^2}{Q^2_0})].
\end{eqnarray}

The three terms behave differently as a function of $Q^2$. Except for
a possible $Q^2$ dependence of the power $n_T$, $F^T_{q\overline{q}}$
does not depend on $Q^2$ as a result of the limited quark $p_T$ in the
aligned configuration. The term $F^L_{q \overline{q}}$ is higher twist
but the power $1/Q^2$ is softened by a logarithmic $Q^2$ factor;
$F^T_{q\overline{q}g}$ grows logarithmically with $Q^2$ similar to the
proton structure function $F_2$ at low $x$.

The coefficients $c_T, c_L, c_g$ as well as the parameters $n^0_T,
n^0_L$ and $x_0, Q^2_0$ were determined from experiment. In the fit
the power $\gamma$ was also considered as a free parameter. Assuming
$Q^2_0 = 1$ GeV$^2$ and $x_0 = 0.0042$ and treating the other
constants as free parameters a good fit ($\chi^2/d.o.f. = 56/47 =
1.2$, statistical errors only) was obtained for the $\xpom
F^{D(3)}_2(\xpom,\beta,Q^2)$ data from this analysis as shown by the
solid curves in Fig.~\ref{f:xf2dfinaltheo}. The fit yielded the
following parameter values: $n^0_T = 0.13 \pm 0.03, \; n^0_L = 0.32
\pm 0.14, \; \gamma = 3.9 \pm 0.9, \; c_T = 0.11 \pm 0.01, \; c_L =
0.12 \pm 0.03, \; c_g = 0.014 \pm 0.002 $; the errors include the
statistical and systematic uncertainties combined in quadrature.

\begin{figure}[ht]
\begin{center}
\epsfig{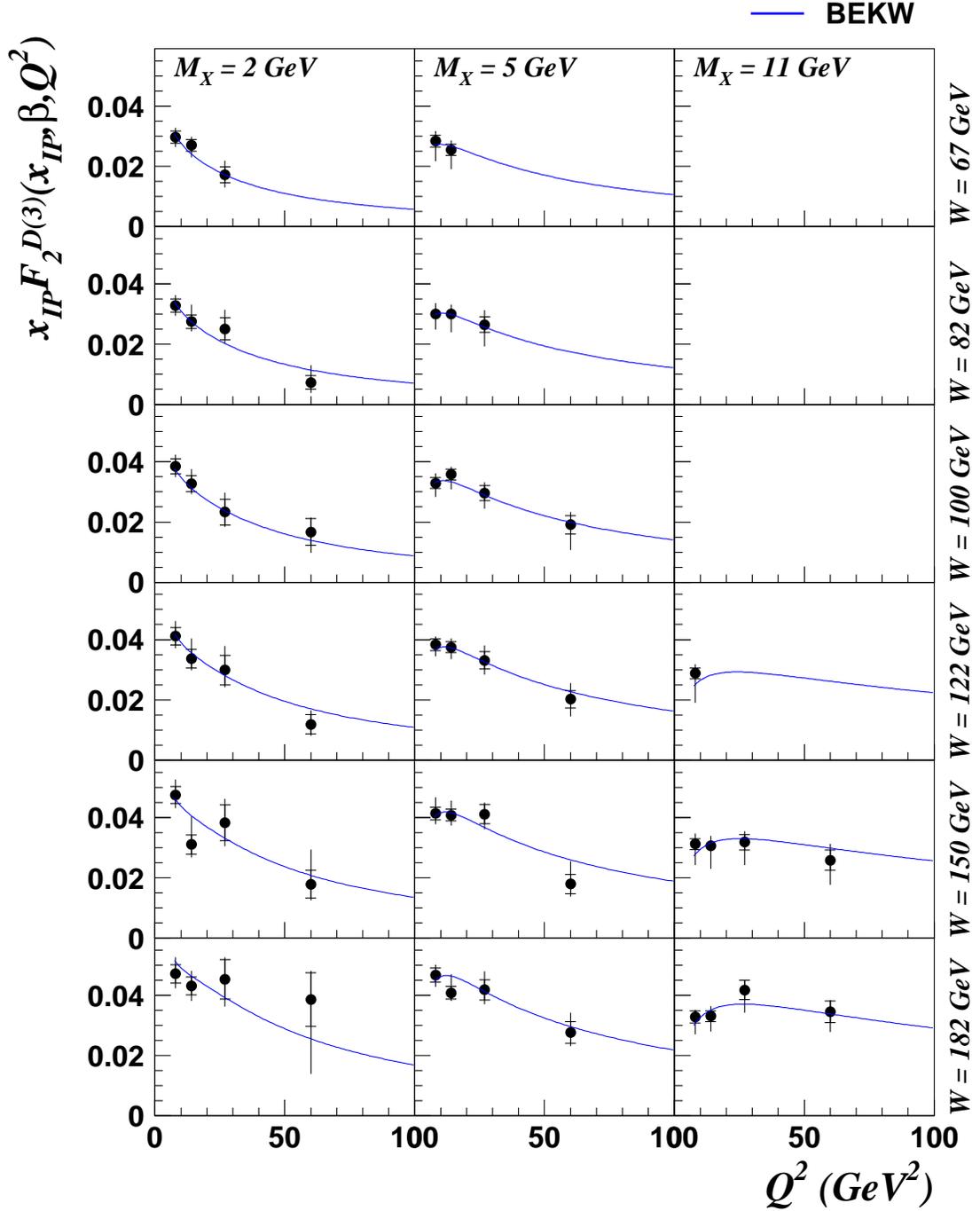}
\end{center}
\vspace{-0.6cm}
\caption[]{The diffractive structure function of the proton 
         for $\gamma^* p \to XN, M_N < 5.5$ GeV, multiplied by
         $\xpom$, $\xpom F^{D(3)}_2(\xpom,\beta,Q^2)$, as a function
         of $Q^2$ from this analysis.  The inner error bars show the
         statistical errors and the full bars the statistical and
         systematic errors added in quadrature.  The curves show the
         fit results obtained with the BEKW model.}
\label{f:xf2dvsq2final}
\end{figure}

The BEKW model also describes the $Q^2$ dependence of $\xpom
F^{D(3)}_2$ as shown by the solid curves in
Fig.~\ref{f:xf2dvsq2final}, and the $\beta$ (and $Q^2$) dependence of
$F^{D(2)}_2$ shown in Fig.~\ref{f:f2dvsbeta}. The value $\gamma = 3.9
\pm 0.9$ is consistent with the prediction of~\cite{Wusthoff},
$\gamma=3$, and somewhat higher than the value $\gamma = 2$ given
in~\cite{Nikzak}.

\begin{figure}[ht]
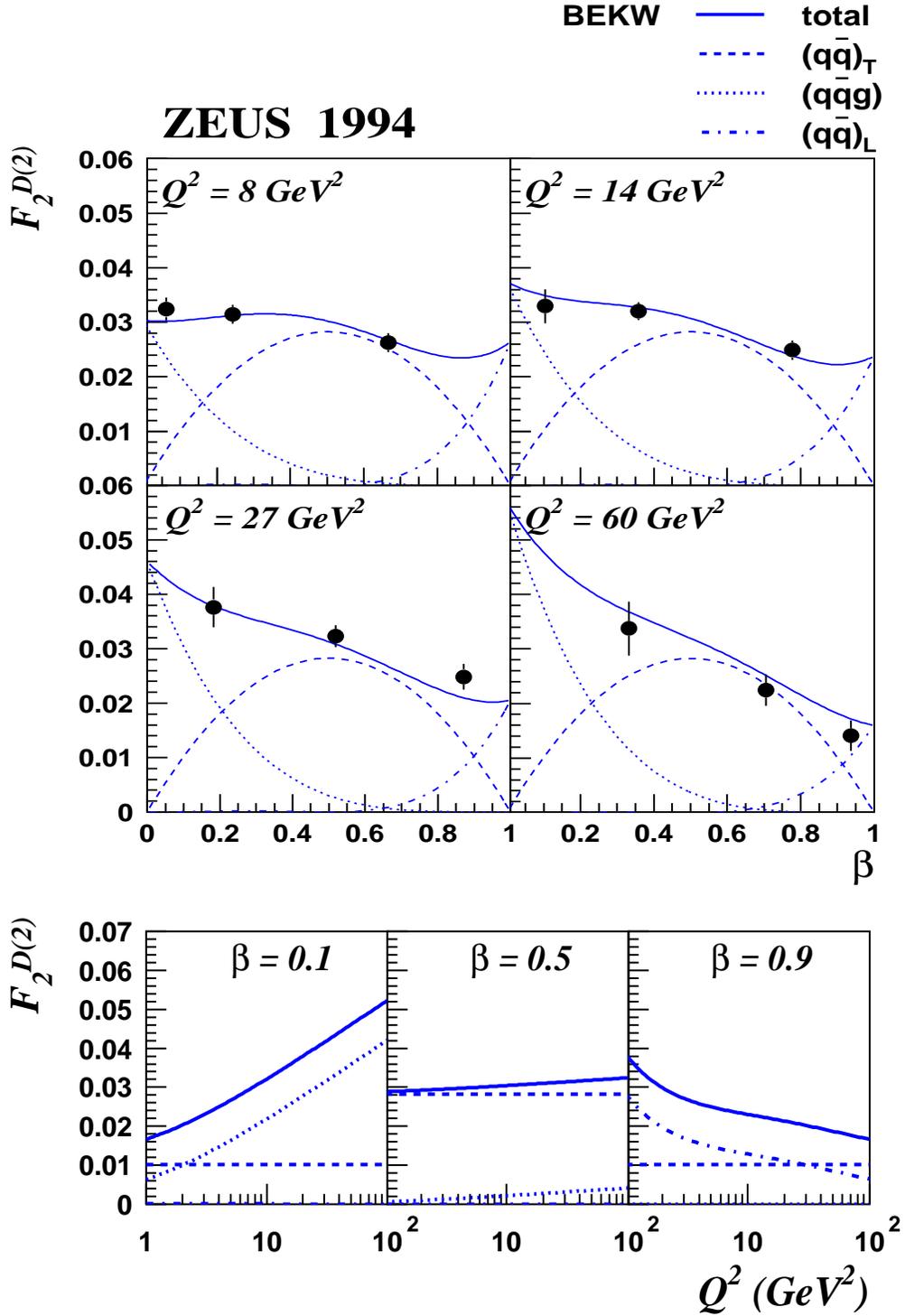

\begin{center}
\epsfig{file=Fig.14a,width=16cm,height=14cm}
\epsfig{file=Fig.14b,bburx=458pt,bbury=170pt,width=16cm,height=6cm}
\end{center}
\vspace{-0.6cm}
\caption[]{Top: The three components $(q\overline{q})_T$, $(q\overline{q}g)$
         and $(q\overline{q})_L$ of the BEKW model building up the
         diffractive structure function of the proton and their sum
         $F^{D(2)}_2(\beta,Q^2)$ as a function of $\beta$ for $Q^2 =
         8, 14, 27$ and $60$ GeV$^2$, as obtained from a fit of the
         BEKW model to the data.  Bottom: the same quantities as a
         function of $Q^2$ for $\beta = 0.1, 0.5$ and $0.9$.}
\label{f:f2d3bw}
\end{figure}

It is instructive to compare the $\beta$ and $Q^2$ dependences of the
three components which build up the diffractive structure function
$F^{D(3)}_2$ in the BEKW model using the results from the
fit. Figure~\ref{f:f2d3bw}(top) shows $c_T F^T_{q\overline{q}}$
(dashed), $c_L F^L_{q\overline{q}}$ (dashed-dotted), $c_g
F^T_{q\overline{q}g}$ (dotted) and their sum $\xpom
F^{D(3)}_2(\xpom,\beta, Q^2)$ at $\xpom = x_0$ (solid curves) as a
function of $\beta$ for $Q^2 = 8, 14, 27, 60$ GeV$^2$. Our data
suggest that for $\beta > 0.2$ the colourless system couples
predominantly to the quarks in the virtual photon. The region $\beta
\ge 0.8$ is dominated by the contributions from longitudinal
photons~\footnote{In determining the diffractive cross section and the
diffractive structure function the term
$[1~-~\frac{y^2}{(1-y)^2+1}\frac{\sigma_L^{di\!f\!f}}{\sigma_L^{di\!f\!f}+
\sigma_T^{di\!f\!f}}]$
has been neglected, see Eq.~\ref{eq:sgp}. If this term is kept for
$\beta > 0.8$ and the BEKW fit is repeated with the assumption
$\sigma_L = \sigma_T$ the changes in the fit parameters are small
compared to their errors.}. The contribution from coupling of the
colourless system to a $q \overline{q} g$ final state becomes
important for $\beta < 0.3$. The last result is in contrast to the H1
observation~\cite{H1epf2d394} that, using a DGLAP NLO fit, the large
$\beta$ region is dominated by the gluon
contribution~\footnote{In~\cite{Bartels98} two possible solutions were
found from fits to the H1 data: one where the gluon term is dominant
at large $\beta$ and one where it is not. The latter had a slightly
larger $\chi^2$ value.}.

Figure~\ref{f:f2d3bw}(bottom) shows the same quantities as a function
of $Q^2$ for $\beta = 0.1, 0.5, 0.9$. The gluon term, which dominates
at $\beta = 0.1$ rises with $Q^2$ while the quark term, which is
important at $\beta = 0.5$ shows no evolution with $Q^2$. The
contribution from longitudinal photons, which is higher twist and
dominates at $\beta = 0.9$, decreases with $Q^2$.

In the BEKW model the $\xpom$-dependence of the quark and gluon
contributions for transverse photons is expected to be close to that
given by the soft pomeron, $n_T \approx 2(\overline{\alphapom}^{soft}
- 1)$. However, perturbative admixtures in the diffractive final state
are expected to have a somewhat stronger energy dependence, leading to
an effective $n_T > 2(\overline{\alphapom}^{soft} - 1)$. The $\xpom$
dependence of the longitudinal contribution is driven by the square of
the proton's gluon momentum density leading to $n_L > n_T$. The fit
results agree with these predictions but the errors are too large for
a definitive statement.

 The same conclusion is reached when separate fits are performed for
 the regions $\beta \ge 0.8$ and $\beta < 0.8$. Assuming $n=n_T = n_L$
 the results are $n(\beta \ge 0.8)~=~0.46 \pm 0.12$ and $n(\beta <
 0.8) = 0.27 \pm 0.03$. It is important to note that already at $Q^2 =
 8$ GeV$^2$, $n_T(Q^2 = 8$~GeV$^2)~=~0.25 \pm~0.04$ which is
 substantially larger than the expectation for soft contributions,
 $n_{soft}~=~0.152^{+0.024}_{-0.018}$, indicating that the transverse
 and gluon components receive sizeable contributions from perturbative
 processes.

In the BH model~\cite{Buchmuller95} the $\xpom$ and $Q^2$ dependences
of the diffractive structure function at small $\xpom$ have been
related to the $x$ and $Q^2$ dependences of the structure function
$F_2$ by assuming that in diffractive DIS a colourless cluster
$\sigma$ is separated from the proton which interacts with the virtual
photon. The probability, $\sigma (\xpom,Q^2)$, for finding such a
cluster in the proton at small $\xpom$ is expected to have an $\xpom$
dependence similar to the $x$ dependence of the quark and gluon
densities in the proton, $g(x,Q^2)$, $q_{sea}(x,Q^2)$, provided $x =
\xpom$. Since diffractive DIS is expected to predominantly produce
configurations where the relative transverse momenta of at least one
pair of partons are small, QCD evolution is suppressed in contrast to
inclusive deep-inelastic scattering. These arguments have led to the
prediction $\xpom F^{D(3)}_2(\xpom,\beta,Q^2) \propto
F_2(x=\xpom,Q^2)/\log_{10}~(Q^2/Q^2_0)$. Here, $Q^2_0 \approx 0.55$
GeV$^2$ was taken from an analysis of the $F_2(x,Q^2)$ data from
HERA~\cite{Buchmuller97}. This relation predicts similar $W$
dependences for the diffractive and total cross sections which is in
agreement with the data presented above. It also predicts different
$Q^2$ dependences for $\xpom F^{D(3)}_2$ and $F_2$.

\begin{figure}[ht]
\begin{center}
\epsfig{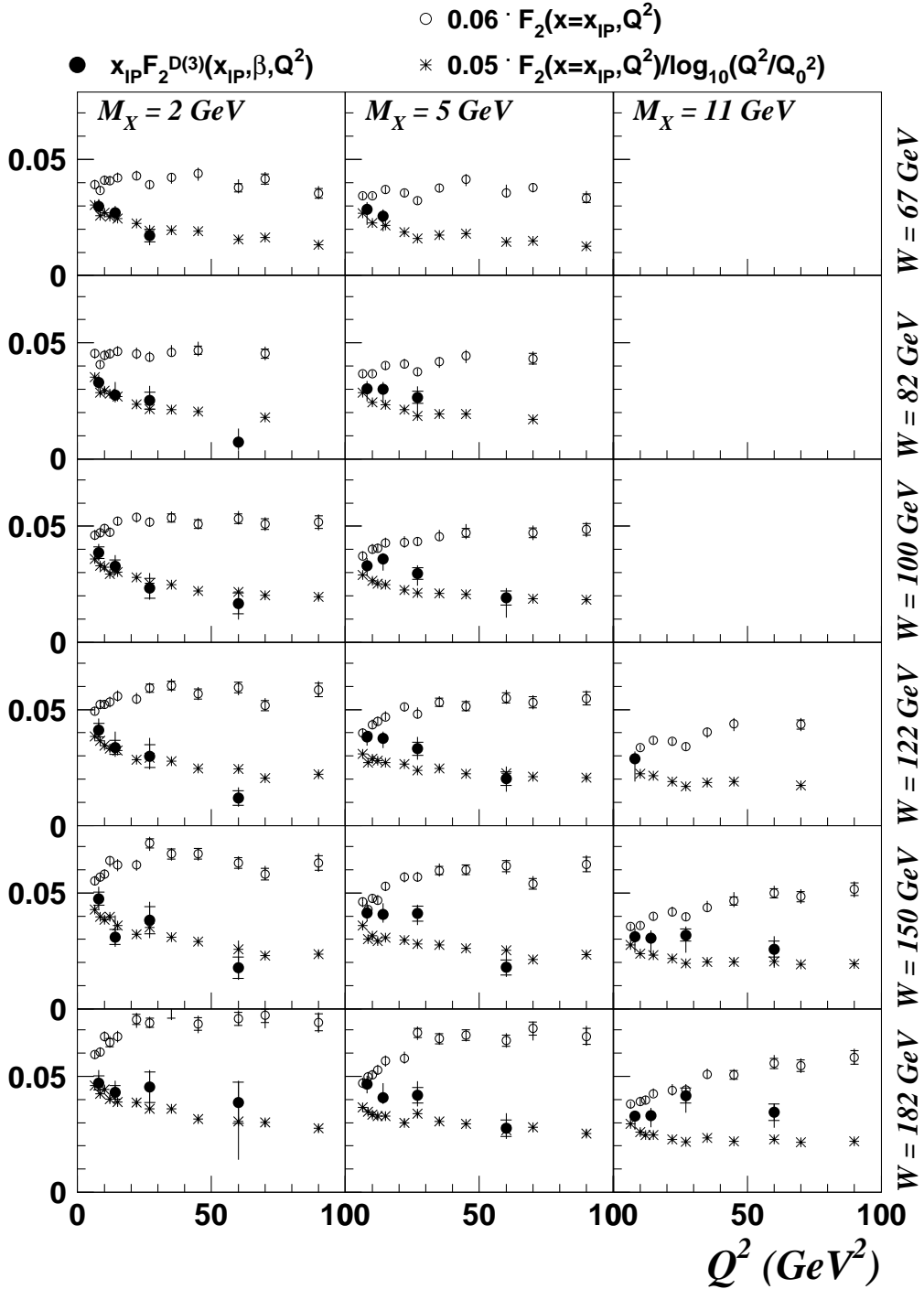}
\end{center}
\vspace{-0.6cm}
\caption[]{The solid points show the diffractive structure function
         for $\gamma^* p \to XN, M_N < 5.5$ GeV, multiplied by
         $\xpom$, $\xpom F^{D(3)}_2(\xpom,\beta,Q^2)$, from this
         analysis, as a function of $Q^2$ for the ($M_X,W$) values
         indicated.  The inner error bars show the statistical errors
         and the full bars the statistical and systematic errors added
         in quadrature.  The open points show the structure function
         $F_2(x=\xpom,Q^2)$ multiplied by 0.06 obtained with the 1994
         data~\protect\cite{Zepf294} as a function of $Q^2$ for the
         ($M_X,Q^2$) values indicated.  The points marked by stars
         show $F_2(x=\xpom,Q^2)/log_{10}(Q^2 / Q^2_0)$ scaled by a
         factor 0.05.  Note that $\xpom$ is known when $M_X,W$ and
         $Q^2$ are given.}
\label{f:xf2d3f2vsq2}
\end{figure}

The $Q^2$ behaviour of the two structure functions is compared in
Fig.~\ref{f:xf2d3f2vsq2} which shows \newline
$\xpom~F^{D(3)}_2(\xpom,\beta,Q^2)$ (solid points) for fixed values of
$M_X$ and $W$ and $F_2(x=\xpom,Q^2)$ (open points). The $F_2(x =
\xpom,Q^2)$ values were obtained from our published
data~\cite{Zepf294} by taking that measurement of $F_2(x,Q^2)$ with
$x$ closest to $\xpom$ and transporting it to $x = \xpom$. For ease of
comparison, $F_2$ has been multiplied by a constant factor of
0.06. The comparison shows that the two quantities have different
evolution with $Q^2$. For $M_X < 7.5$ GeV, $\xpom F^{D(3)}_2$
decreases with $Q^2$ while the structure function $F_2(x,Q^2)$
gradually rises with $Q^2$. In Fig.~\ref{f:xf2d3f2vsq2} $\xpom
F^{D(3)}_2$ is also compared with
$F_2(x=\xpom,Q^2)/\log_{10}(Q^2/Q^2_0)$ (points marked as stars) as
suggested by~\cite{Buchmuller95}. Here, the $F_2$ values were
multiplied by a factor of 0.05 as obtained from a fit to the $\xpom
F^{D(3)}_2$ for $M_X = 2$ GeV. The $Q^2$ evolution of the data at low
$M_X$ ($M_X < 3$ GeV) is well described by this model. At larger $M_X$
values there is a tendency for the data to lie above the BH
prediction. We note that in the BEKW model this is understood as
resulting from the logarithmic growth of the ($q\overline{q}g$)
contribution with $Q^2$.

\section{Summary and conclusion}

The DIS diffractive cross section $d\sigma^{di\!f\!f}_{\gamma^* p \to
XN}/dM_X$, has been measured for $M_N < 5.5$ GeV, $M_X < 15$ GeV, $60
< W < 200$ GeV and $7 < Q^2 < 140$ GeV$^2$. For fixed $Q^2$ the
diffractive cross section rises rapidly with $W$. A fit of the $W$
dependence by the form $d\sigma^{di\!f\!f}_{\gamma^*p \to
XN}(M_X,W,Q^2)/dM_X \propto W^{a^{diff}}$ yielded $a^{diff} = 0.507
\pm 0.034(stat)^{+0.155}_{-0.046} (syst)$ which corresponds to a
$t$-averaged pomeron trajectory of $\overline{\alphapom} = 1.127 \pm
0.009 (stat)^{+0.039}_{-0.012} (syst)$. The rise is faster than
expected in Regge models using the intercept of the pomeron trajectory
extracted from hadron-hadron scattering. The $W$ dependence of the
diffractive cross section, contrary to na\"{\i}ve expectations, is the
same as that of the total virtual photon proton cross section. The
diffractive contribution to the total cross section for $M_X < 15$
GeV, $M_N < 5.5$ GeV and $ 134 < W < 200$ GeV amounts to
$13.2^{+0.5}_{-0.7}\%$ at $Q^2 = 8$ GeV$^2$ decreasing to
$3.7^{+0.3}_{-0.4}\%$ at $Q^2 = 60$ GeV$^2$.

The analysis of the data in terms of the diffractive structure
function $F^{D(3)}_2(\xpom,\beta,Q^2)$ of the proton shows that $\xpom
F^{D(3)}_2$ rises as $\xpom \to 0$. The data are consistent with the
assumption that the diffractive structure function $F^{D(3)}_2$
factorizes into a term depending only on $\xpom$ and a structure
function $F^{D(2)}_2$ which depends on ($\beta,Q^2$). The rise of
$\xpom F^{D(3)}_2$ with $\xpom$ can be described as $\xpom F^{D(3)}_2
\propto (1/\xpom)^n$ with $n = 0.253 \pm 0.017
(stat)^{+0.077}_{-0.023}(syst)$. The data are also consistent with
models which break factorization. The rise of $F^{D(3)}_2$ reflects
the rise of $d\sigma^{di\!f\!f}_{\gamma^* p \to XN}/dM_X$ with
$W$. For fixed $M_X < 7.5$ GeV and fixed $W$, $\xpom F^{D(3)}_2$
decreases slowly with $Q^2$ while for $M_X = 11$ GeV it is
approximately constant.

The data have been compared with several partonic models of
diffraction. Good agreement with the data can be achieved. The models
provide a first glimpse of how the different components may build up
the diffractive structure function.  The $Q^2$ behaviour of $\xpom
F^{D(3)}_2(\xpom,\beta,Q^2)$ is different from that of the proton
structure function $F_2(x,Q^2)$, taken at $x = \xpom$, which rises
gradually with $Q^2$. It is in broad agreement with the BH conjecture
that $\xpom F^{D(3)}_2(\xpom,\beta,Q^2) \propto
F_2(x=\xpom,Q^2)/\log_{10}(Q^2 / Q^2_0)$ where $Q^2_0 = 0.55$
GeV$^2$. Using the BEKW model at medium $\beta$ the main contribution
comes from transverse photons coupling to a $q\overline{q}$
system. The region $\beta < 0.2$ is dominated by $q\overline{q}g$
contributions. Longitudinal photons coupling to a $q\overline{q}$
system account for most of the data at $\beta > 0.8$. The transverse
photon $q\overline{q}$ contribution, which is dominant, is of leading
twist and has no substantial evolution with $Q^2$.

The leading twist behaviour and the strong rise of $\xpom F^{D(3)}_2$
as $\xpom \to 0$ suggest a partonic process as a major production
mechanism for diffractive scattering in DIS.

\newpage

\begin{table}[hbt]

\scriptsize
\caption{Cross section for diffractive scattering via $\gamma^*p \to X N$, where
$N$ is the proton or dissociated nucleonic system with mass $M_N <
5.5$ GeV as a function of $M_X,Q^2$ and $W$. The statistical and
systematic errors are given. The overall normalization uncertainty of
$\pm 2\%$ is not included.  }
\centering
\vspace{0.5cm}
\begin{tabular}{|r|r|r|rrr|}
\hline
$M_X$ & $Q^2$     & $W$   & $d\sigma^{diff}_{\gamma^*p \to X N}/dM_X$ & $\pm$
 stat & $\pm$ syst \\
(GeV) & (GeV$^2$) & (GeV) &  & (nb/GeV) &                          \\
\hline
 2.0 &   8.0 &  66.7 &  138.6 &  $\pm$  9.5 & $^{+11.8}_{-11.6}$  \\
 2.0 &   8.0 &  81.8 &  153.8 &  $\pm$ 10.1 & $^{+11.8}_{-12.2}$  \\
 2.0 &   8.0 &  99.8 &  180.0 &  $\pm$ 11.7 & $^{+13.6}_{-11.7}$  \\
 2.0 &   8.0 & 122.1 &  192.3 &  $\pm$ 13.4 & $^{+19.6}_{-13.3}$  \\
 2.0 &   8.0 & 148.6 &  221.9 &  $\pm$ 13.2 & $^{+20.2}_{-16.5}$  \\
 2.0 &   8.0 & 181.5 &  219.9 &  $\pm$ 13.9 & $^{+22.2}_{-17.5}$  \\
\hline
 5.0 &   8.0 &  66.9 &  120.3 &  $\pm$  7.9 & $^{+11.5}_{-27.1}$  \\
 5.0 &   8.0 &  81.6 &  127.9 &  $\pm$  7.0 & $^{+13.1}_{-21.2}$  \\
 5.0 &   8.0 &  99.9 &  139.8 &  $\pm$  7.5 & $^{+12.1}_{-17.7}$  \\
 5.0 &   8.0 & 121.9 &  162.8 &  $\pm$  8.2 & $^{+ 8.0}_{-14.0}$  \\
 5.0 &   8.0 & 148.9 &  175.4 &  $\pm$  8.8 & $^{+20.6}_{-12.4}$  \\
 5.0 &   8.0 & 181.7 &  198.4 &  $\pm$ 10.0 & $^{+11.0}_{-13.6}$  \\
\hline
11.0 &   8.0 & 121.7 &   68.9 &  $\pm$  4.2 & $^{+ 5.9}_{-22.9}$  \\
11.0 &   8.0 & 149.1 &   74.5 &  $\pm$  4.4 & $^{+ 7.0}_{-16.1}$  \\
11.0 &   8.0 & 181.6 &   78.4 &  $\pm$  4.8 & $^{+ 2.7}_{-12.9}$  \\
\hline
 2.0 &  14.0 &  67.2 &   48.0 &  $\pm$  3.6 & $^{+ 3.7}_{- 6.1}$  \\
 2.0 &  14.0 &  81.5 &   48.9 &  $\pm$  4.1 & $^{+ 9.2}_{- 4.2}$  \\
 2.0 &  14.0 & 100.0 &   58.1 &  $\pm$  4.8 & $^{+ 7.2}_{- 4.0}$  \\
 2.0 &  14.0 & 121.6 &   60.0 &  $\pm$  5.5 & $^{+10.6}_{- 4.4}$  \\
 2.0 &  14.0 & 148.5 &   55.2 &  $\pm$  5.7 & $^{+15.6}_{- 5.0}$  \\
 2.0 &  14.0 & 181.8 &   76.7 &  $\pm$  5.3 & $^{+ 6.9}_{- 7.1}$  \\
\hline
 5.0 &  14.0 &  67.0 &   52.3 &  $\pm$  3.7 & $^{+ 5.2}_{-12.7}$  \\
 5.0 &  14.0 &  81.7 &   61.4 &  $\pm$  3.2 & $^{+ 5.5}_{-11.9}$  \\
 5.0 &  14.0 &  99.9 &   73.5 &  $\pm$  3.7 & $^{+ 3.8}_{- 9.5}$  \\
 5.0 &  14.0 & 121.7 &   76.9 &  $\pm$  3.8 & $^{+ 4.7}_{- 7.5}$  \\
 5.0 &  14.0 & 148.9 &   83.7 &  $\pm$  4.1 & $^{+ 9.1}_{- 5.9}$  \\
 5.0 &  14.0 & 182.0 &   83.8 &  $\pm$  4.3 & $^{+12.0}_{- 3.3}$  \\
\hline
11.0 &  14.0 & 149.1 &   40.0 &  $\pm$  2.2 & $^{+ 3.7}_{- 9.7}$  \\
11.0 &  14.0 & 181.6 &   43.2 &  $\pm$  2.4 & $^{+ 3.7}_{- 6.2}$  \\
\hline
 2.0 &  27.0 &  67.2 &    9.1 &  $\pm$  1.4 & $^{+ 2.0}_{- 1.8}$  \\
 2.0 &  27.0 &  82.2 &   13.4 &  $\pm$  2.0 & $^{+ 2.8}_{- 1.7}$  \\
 2.0 &  27.0 &  99.4 &   12.5 &  $\pm$  2.3 & $^{+ 2.6}_{- 1.3}$  \\
 2.0 &  27.0 & 121.4 &   16.0 &  $\pm$  2.6 & $^{+ 3.4}_{- 1.6}$  \\
 2.0 &  27.0 & 148.9 &   20.5 &  $\pm$  3.2 & $^{+ 2.8}_{- 2.7}$  \\
 2.0 &  27.0 & 182.0 &   24.3 &  $\pm$  3.5 & $^{+ 1.7}_{- 3.4}$  \\
\hline
 5.0 &  27.0 &  81.7 &   21.1 &  $\pm$  2.0 & $^{+ 3.2}_{- 5.4}$  \\
 5.0 &  27.0 &  99.5 &   23.6 &  $\pm$  2.0 & $^{+ 2.0}_{- 3.6}$  \\
 5.0 &  27.0 & 121.8 &   26.4 &  $\pm$  2.3 & $^{+ 3.2}_{- 3.0}$  \\
 5.0 &  27.0 & 149.2 &   32.8 &  $\pm$  2.6 & $^{+ 1.6}_{- 3.1}$  \\
 5.0 &  27.0 & 181.1 &   33.4 &  $\pm$  2.7 & $^{+ 4.1}_{- 2.7}$  \\
\hline
11.0 &  27.0 & 148.9 &   19.6 &  $\pm$  1.6 & $^{+ 1.6}_{- 4.4}$  \\
11.0 &  27.0 & 181.5 &   25.8 &  $\pm$  2.0 & $^{+ 0.6}_{- 4.1}$  \\
\hline
 2.0 &  60.0 &  81.2 &    0.8 &  $\pm$  0.3 & $^{+ 0.6}_{- 0.3}$  \\
 2.0 &  60.0 & 101.1 &    1.9 &  $\pm$  0.5 & $^{+ 0.2}_{- 0.6}$  \\
 2.0 &  60.0 & 122.5 &    1.4 &  $\pm$  0.4 & $^{+ 0.4}_{- 0.2}$  \\
 2.0 &  60.0 & 148.8 &    2.1 &  $\pm$  0.5 & $^{+ 1.2}_{- 0.3}$  \\
 2.0 &  60.0 & 180.2 &    4.5 &  $\pm$  1.0 & $^{+ 0.4}_{- 2.7}$  \\
\hline
 5.0 &  60.0 &  99.4 &    4.2 &  $\pm$  0.7 & $^{+ 0.7}_{- 1.7}$  \\
 5.0 &  60.0 & 122.5 &    4.4 &  $\pm$  0.6 & $^{+ 1.0}_{- 1.1}$  \\
 5.0 &  60.0 & 149.2 &    3.9 &  $\pm$  0.7 & $^{+ 1.5}_{- 0.6}$  \\
 5.0 &  60.0 & 182.2 &    6.1 &  $\pm$  0.8 & $^{+ 1.2}_{- 0.6}$  \\
\hline
11.0 &  60.0 & 148.7 &    5.8 &  $\pm$  0.8 & $^{+ 1.0}_{- 1.7}$  \\
11.0 &  60.0 & 181.6 &    7.8 &  $\pm$  0.8 & $^{+ 0.3}_{- 1.3}$  \\
\hline
\end{tabular}
\label{t:diffcrosstab}
\end{table}
\newpage

\begin{table}[hbt]
\scriptsize
\caption{The diffractive structure function multiplied by $\xpom$,
$\xpom F^{D(3)}_2(\xpom,\beta,Q^2)$, for diffractive scattering via
$\gamma^*p \to X N$, where $N$ is a nucleonic system with mass $M_N <
5.5$ GeV as a function of $\xpom$, $\beta$ and $Q^2$. The statistical
and systematic errors are given. The overall normalization uncertainty
of $\pm 2\%$ is not included. }  \centering
\vspace{0.5cm}
\begin{tabular}{|r|r|r|rrr|}
\hline
$\xpom$ & $\beta$     & $Q^2$   & $\xpom F^{D(3)}_2$ & $\pm$ stat & $\pm$ syst \\
           &            & (GeV$^2$) &                     &            &          \\
\hline
\hline
 0.00269 & 0.667 &   8.0 & 0.0297 & $\pm $ 0.0020 & $ ^{+0.0025}_{-0.0025}$ \\
 0.00179 & 0.667 &   8.0 & 0.0329 & $\pm $ 0.0022 & $ ^{+0.0025}_{-0.0026}$ \\
 0.00120 & 0.667 &   8.0 & 0.0385 & $\pm $ 0.0025 & $ ^{+0.0029}_{-0.0025}$ \\
 0.00081 & 0.667 &   8.0 & 0.0411 & $\pm $ 0.0029 & $ ^{+0.0042}_{-0.0028}$ \\
 0.00054 & 0.667 &   8.0 & 0.0475 & $\pm $ 0.0028 & $ ^{+0.0043}_{-0.0035}$ \\
 0.00036 & 0.667 &   8.0 & 0.0471 & $\pm $ 0.0030 & $ ^{+0.0047}_{-0.0037}$ \\
 0.00735 & 0.242 &   8.0 & 0.0284 & $\pm $ 0.0019 & $ ^{+0.0027}_{-0.0064}$ \\
 0.00495 & 0.242 &   8.0 & 0.0301 & $\pm $ 0.0016 & $ ^{+0.0031}_{-0.0050}$ \\
 0.00330 & 0.242 &   8.0 & 0.0329 & $\pm $ 0.0018 & $ ^{+0.0028}_{-0.0042}$ \\
 0.00222 & 0.242 &   8.0 & 0.0383 & $\pm $ 0.0019 & $ ^{+0.0019}_{-0.0033}$ \\
 0.00149 & 0.242 &   8.0 & 0.0413 & $\pm $ 0.0021 & $ ^{+0.0049}_{-0.0029}$ \\
 0.00100 & 0.242 &   8.0 & 0.0467 & $\pm $ 0.0023 & $ ^{+0.0026}_{-0.0032}$ \\
 0.00871 & 0.062 &   8.0 & 0.0288 & $\pm $ 0.0018 & $ ^{+0.0025}_{-0.0096}$ \\
 0.00580 & 0.062 &   8.0 & 0.0312 & $\pm $ 0.0018 & $ ^{+0.0029}_{-0.0067}$ \\
 0.00391 & 0.062 &   8.0 & 0.0328 & $\pm $ 0.0020 & $ ^{+0.0011}_{-0.0054}$ \\
\hline
 0.00398 & 0.778 &  14.0 & 0.0270 & $\pm $ 0.0020 & $ ^{+0.0021}_{-0.0034}$ \\
 0.00270 & 0.778 &  14.0 & 0.0275 & $\pm $ 0.0023 & $ ^{+0.0052}_{-0.0023}$ \\
 0.00180 & 0.778 &  14.0 & 0.0327 & $\pm $ 0.0027 & $ ^{+0.0041}_{-0.0022}$ \\
 0.00122 & 0.778 &  14.0 & 0.0337 & $\pm $ 0.0031 & $ ^{+0.0060}_{-0.0025}$ \\
 0.00082 & 0.778 &  14.0 & 0.0310 & $\pm $ 0.0032 & $ ^{+0.0087}_{-0.0028}$ \\
 0.00054 & 0.778 &  14.0 & 0.0431 & $\pm $ 0.0030 & $ ^{+0.0039}_{-0.0040}$ \\
 0.00866 & 0.359 &  14.0 & 0.0255 & $\pm $ 0.0018 & $ ^{+0.0025}_{-0.0062}$ \\
 0.00583 & 0.359 &  14.0 & 0.0300 & $\pm $ 0.0016 & $ ^{+0.0027}_{-0.0058}$ \\
 0.00390 & 0.359 &  14.0 & 0.0358 & $\pm $ 0.0018 & $ ^{+0.0018}_{-0.0046}$ \\
 0.00263 & 0.359 &  14.0 & 0.0375 & $\pm $ 0.0018 & $ ^{+0.0023}_{-0.0036}$ \\
 0.00176 & 0.359 &  14.0 & 0.0408 & $\pm $ 0.0020 & $ ^{+0.0044}_{-0.0029}$ \\
 0.00118 & 0.359 &  14.0 & 0.0408 & $\pm $ 0.0021 & $ ^{+0.0059}_{-0.0016}$ \\
 0.00607 & 0.104 &  14.0 & 0.0306 & $\pm $ 0.0017 & $ ^{+0.0028}_{-0.0074}$ \\
 0.00409 & 0.104 &  14.0 & 0.0331 & $\pm $ 0.0018 & $ ^{+0.0028}_{-0.0048}$ \\
\hline
 0.00682 & 0.871 &  27.0 & 0.0172 & $\pm $ 0.0027 & $ ^{+0.0038}_{-0.0033}$ \\
 0.00457 & 0.871 &  27.0 & 0.0251 & $\pm $ 0.0037 & $ ^{+0.0052}_{-0.0032}$ \\
 0.00313 & 0.871 &  27.0 & 0.0233 & $\pm $ 0.0043 & $ ^{+0.0048}_{-0.0024}$ \\
 0.00210 & 0.871 &  27.0 & 0.0299 & $\pm $ 0.0049 & $ ^{+0.0063}_{-0.0031}$ \\
 0.00140 & 0.871 &  27.0 & 0.0382 & $\pm $ 0.0059 & $ ^{+0.0053}_{-0.0050}$ \\
 0.00093 & 0.871 &  27.0 & 0.0453 & $\pm $ 0.0066 & $ ^{+0.0031}_{-0.0063}$ \\
 0.00776 & 0.519 &  27.0 & 0.0265 & $\pm $ 0.0026 & $ ^{+0.0040}_{-0.0068}$ \\
 0.00524 & 0.519 &  27.0 & 0.0296 & $\pm $ 0.0025 & $ ^{+0.0025}_{-0.0045}$ \\
 0.00350 & 0.519 &  27.0 & 0.0331 & $\pm $ 0.0029 & $ ^{+0.0041}_{-0.0037}$ \\
 0.00233 & 0.519 &  27.0 & 0.0411 & $\pm $ 0.0032 & $ ^{+0.0020}_{-0.0039}$ \\
 0.00158 & 0.519 &  27.0 & 0.0418 & $\pm $ 0.0034 & $ ^{+0.0051}_{-0.0034}$ \\
 0.00667 & 0.182 &  27.0 & 0.0318 & $\pm $ 0.0026 & $ ^{+0.0026}_{-0.0071}$ \\
 0.00449 & 0.182 &  27.0 & 0.0417 & $\pm $ 0.0032 & $ ^{+0.0010}_{-0.0067}$ \\
\hline
 0.00961 & 0.938 &  60.0 & 0.0073 & $\pm $ 0.0023 & $ ^{+0.0052}_{-0.0026}$ \\
 0.00622 & 0.938 &  60.0 & 0.0167 & $\pm $ 0.0045 & $ ^{+0.0017}_{-0.0053}$ \\
 0.00425 & 0.938 &  60.0 & 0.0119 & $\pm $ 0.0032 & $ ^{+0.0033}_{-0.0021}$ \\
 0.00288 & 0.938 &  60.0 & 0.0178 & $\pm $ 0.0046 & $ ^{+0.0106}_{-0.0027}$ \\
 0.00197 & 0.938 &  60.0 & 0.0386 & $\pm $ 0.0089 & $ ^{+0.0035}_{-0.0230}$ \\
 0.00856 & 0.706 &  60.0 & 0.0191 & $\pm $ 0.0030 & $ ^{+0.0031}_{-0.0079}$ \\
 0.00564 & 0.706 &  60.0 & 0.0202 & $\pm $ 0.0029 & $ ^{+0.0045}_{-0.0049}$ \\
 0.00381 & 0.706 &  60.0 & 0.0179 & $\pm $ 0.0032 & $ ^{+0.0068}_{-0.0026}$ \\
 0.00255 & 0.706 &  60.0 & 0.0276 & $\pm $ 0.0036 & $ ^{+0.0055}_{-0.0028}$ \\
 0.00817 & 0.331 &  60.0 & 0.0258 & $\pm $ 0.0034 & $ ^{+0.0043}_{-0.0074}$ \\
 0.00548 & 0.331 &  60.0 & 0.0345 & $\pm $ 0.0036 & $ ^{+0.0014}_{-0.0057}$ \\
\hline
\end{tabular}
\label{t:xf2d3tab}
\end{table}

\clearpage

\clearpage

\newpage

\clearpage

\clearpage

\clearpage

\newpage

\newpage
\end{document}